\documentclass[aps,prc,twocolumn,showpacs,nofootinbib,superscriptaddress]{revtex4} %  floatfix,,nofootinbib
\usepackage{graphicx,longtable} %,epsfig,
\usepackage{dcolumn}
\usepackage{mathptmx}
\usepackage{amsmath}
\usepackage[pdftex]{hyperref}
\usepackage{wasysym}
\usepackage[sort&compress]{natbib}

\newcommand{\beqy}{\begin{eqnarray}}
\newcommand{\eeqy}{\end{eqnarray}}
\newcommand{\bmlet}{\begin{subequations}}
\newcommand{\emlet}{\end{subequations}}
\newcounter{saveeqn}

\def\gsimeq{\,\,\raise0.14em\hbox{$>$}\kern-0.76em\lower0.28em\hbox  
{$\sim$}\,\,}  
\def\lsimeq{\,\,\raise0.14em\hbox{$<$}\kern-0.76em\lower0.28em\hbox  
{$\sim$}\,\,}  

\begin{document}

\title{Primary $\gamma$-ray spectra in $^{44}$Ti of astrophysical interest}

\author{A.~C.~Larsen}
\email{a.c.larsen@fys.uio.no}
\affiliation{Department of Physics, University of Oslo, N-0316 Oslo, Norway}
\author{S.~Goriely}
\affiliation{Institut d'Astronomie et d'Astrophysique, Universit\'e Libre de Bruxelles, CP 226,  1050 Brussels, Belgium}
%\email{sgoriely@astro.ulb.ac.be}
\author{A.~B\"{u}rger}
\affiliation{Department of Physics, University of Oslo, N-0316 Oslo, Norway}
\author{M.~Guttormsen}
\affiliation{Department of Physics, University of Oslo, N-0316 Oslo, Norway}
\author{A.~G\"{o}rgen}
\affiliation{Dapnia/SPhN, CEA-Saclay, France}
\affiliation{Department of Physics, University of Oslo, N-0316 Oslo, Norway}
\author{S.~Harissopulos}
\affiliation{Institute of Nuclear Physics, NCSR "Demokritos", 153.10 Aghia Paraskevi, Athens, Greece}
\author{M.~Kmiecik}
\affiliation{Institute of Nuclear Physics PAN, Krak\'{o}w, Poland}
\author{T.~Konstantinopoulos}
\affiliation{Institute of Nuclear Physics, NCSR "Demokritos", 153.10 Aghia Paraskevi, Athens, Greece}
%\author{M.~Krti\u{c}ka}
%\affiliation{Institute of Particle and Nuclear Physics, Charles University, Prague, Czech Republic}
\author{A.~Lagoyannis}
\affiliation{Institute of Nuclear Physics, NCSR "Demokritos", 153.10 Aghia Paraskevi, Athens, Greece}
\author{T.~L\"{o}nnroth}
\affiliation{Department of Physics, \AA bo Akademi University, FIN-20500 \AA bo, Finland}
\author{K.~Mazurek}
\affiliation{Institute of Nuclear Physics PAN, Krak\'{o}w, Poland}
\author{M.~Norrby}
\affiliation{Department of Physics, \AA bo Akademi University, FIN-20500 \AA bo, Finland}
\author{H.~T.~Nyhus}
\affiliation{Department of Physics, University of Oslo, N-0316 Oslo, Norway}
\author{G.~Perdikakis\footnote{Current address: 
	National Superconducting Cyclotron Laboratory, Michigan State University, East Lansing, MI 48824-1321, USA.}}
\affiliation{Institute of Nuclear Physics, NCSR "Demokritos", 153.10 Aghia Paraskevi, Athens, Greece}
\author{A.~Schiller}
\affiliation{Department of Physics and Astronomy, Ohio University, Athens, Ohio 45701, USA}
\author{S.~Siem}
\affiliation{Department of Physics, University of Oslo, N-0316 Oslo, Norway}
\author{A.~Spyrou\footnotemark[\value{footnote}]}
\affiliation{Institute of Nuclear Physics, NCSR "Demokritos", 153.10 Aghia Paraskevi, Athens, Greece}
\altaffiliation{National Superconducting Cyclotron Laboratory, Michigan State University, East Lansing, MI 48824, USA}
\author{N.~U.~H.~Syed}
\affiliation{Department of Physics, University of Oslo, N-0316 Oslo, Norway}
\author{H.~K.~Toft}
\affiliation{Department of Physics, University of Oslo, N-0316 Oslo, Norway}
\author{G.~M.~Tveten}
\affiliation{Department of Physics, University of Oslo, N-0316 Oslo, Norway}
\author{A.~Voinov}
\affiliation{Department of Physics and Astronomy, Ohio University, Athens, Ohio 45701, USA}

\date{\today}

\begin{abstract}
Primary $\gamma$-ray spectra for a wide excitation-energy range 
have been extracted for $^{44}$Ti
from particle-$\gamma$ coincidence data of the 
$^{46}$Ti($p,t\gamma$)$^{44}$Ti reaction. 
These spectra reveal information on 
%the functional form of
%the level density and radiative strength function in $^{44}$Ti. 
the $\gamma$-decay pattern of the nucleus, and may be used to extract 
the level density and radiative strength function applying the Oslo method.

%The level density displays the same 
%features as the known levels from spectroscopy data at low excitation energies, giving confidence that the 
%method gives reliable results also for the very special two-neutron pick-up reaction. 
%In general, the radiative strength function resembles the tail of the giant dipole resonance.
%The data seem to be compatible with a weak enhancement in the radiative strength 
%at low $\gamma$-ray energies ($E_{\gamma} < 3$ MeV). 
Models of the level density and radiative strength function
are used as 
input for cross-section calculations
of the $^{40}$Ca($\alpha,\gamma$)$^{44}$Ti reaction. 
Acceptable models should reproduce data on the 
$^{40}$Ca($\alpha,\gamma$)$^{44}$Ti reaction cross section as well as the measured primary $\gamma$-ray spectra. 
This is only achieved when a coherent normalization of the slope of the level density and radiative strength function
is performed. Thus, the overall shape of the experimental primary $\gamma$-ray spectra  puts a constraint
on the input models for the rate calculations. 

\end{abstract}

\pacs{21.10.Ma, 25.20.Lj, 27.40.+z, 25.40.Hs}
% PACS Numbers: 21.10.Ma (level density), 
%               21.10.-k (Properties of nuclei; nuclear energy levels), 
%               21.60.Jz (HF & QRPA)
% 		25.20.Lj (Photoproduction reactions)
%		24.30.Gd Other resonances
%		25.55.Hp	3He transfer reactions
%		27.40.+z	39 ² A ² 58
%		27.50.+e	59 ² A ² 89
%		25.40.Hs	Transfer reactions, nucleon-induced
% From Voinov's PRL: 25.40.Lw, 25.20.Lj, 25.55.Hp, 27.40.+z

%$^*$ Current address: National Superconducting Cyclotron Laboratory, Michigan State University, East Lansing, MI 48824, USA

\maketitle

%\section{NOTES}
%\begin{itemize}
%\item{ clean up the section with primary spectra -- more inputs with SLO model. 
%	Probably also needs to update the RSF models, to include the SLO model there. 
%	Also some discussion of the RSF models.}
%\item{ the last section before the summary needs a thorough revision. separate the calculations and the comparison of data.
%	The inputs are now probably 8, update these.
%	Put in subsections: impact of $f_{\mathrm{iso}}$ and $\Gamma_\gamma$ (uncertainties in x-section calcs), 
%	role of RSF model, role of NLD model, and finally comparison with data.}
%\end{itemize}

\section{Introduction}
\label{sec:int}
The titanium isotope $^{44}$Ti is of great astrophysical interest, since it is believed to be 
produced in the inner regions of core-collapse supernovae and in the normal freeze-out of Si burning layers of thermonuclear
supernovae~\cite{vockenhuber07}, with a large variation of 
yields depending on their type~\cite{renaud06}. The determination of the $^{44}$Ti yield might reveal information on 
the complex explosion conditions. 
The production yield of $^{44}$Ti directly determines the abundance of the stable $^{44}$Ca,
and also influences the $^{48}$Ti abundance through the feeding of $^{48}$Cr on the 
$\alpha$ chain. The theoretical prediction of the $^{44}$Ti production in 
core-collapse supernovae is sensitive to the chosen reaction network and the adopted nuclear reaction rates.

Cassiopeia A (Cas A) is the youngest known Galactic supernova remnant and is, at present, the only one
from which $\gamma$-rays from the $^{44}$Ti decay chain ($^{44}$Ti $\rightarrow$ $^{44}$Sc 
$\rightarrow$ $^{44}$Ca) have been unambiguously detected~\cite{renaud06}. The discovery of 
the 1157-keV $\gamma$-ray line from $^{44}$Ca was reported by Iyudin \textit{et al.}~\cite{iyudin},
while measurements of the 67.9-keV and 78.4-keV lines from $^{44}$Sc were presented by 
Renaud \textit{et al.}~\cite{renaud06}. From the combined $\gamma$-ray flux, the half-life of
$^{44}$Ti, and the distance and age of the remnant, an initial synthesized $^{44}$Ti
mass of 1.6$^{+0.6}_{-0.3} \times 10^{-4}$ $M_{\astrosun}$ was deduced. This is thought to be unusually	
large, a factor of $2-10$ more than what is typically obtained by current models (see, e.g.,~\cite{woosley95,thiele96}). 

The main production reaction of $^{44}$Ti is the $^{40}$Ca($\alpha,\gamma$)$^{44}$Ti reaction channel
with a $Q$-value of 5.127 MeV;
the cross section for this reaction is very important to estimate the $^{44}$Ti yield. Recent 
cross-section measurements on this reaction~\cite{Nassar06} have lead to an increase of the associated 
astrophysical rate, giving a factor of $\sim 2$ more in the predicted yield of $^{44}$Ti. Thus, the theoretical
models become more compatible with the Cas A data; however, this would make the problem of "young, missing, 
and hidden" Galactic supernova remnants even more serious: supernovae that should have occured after Cas A
are still not detected by means of $\gamma$-ray emission from the $^{44}$Ti decay chain. It is still an open 
question whether the Cas A is a peculiar case (asymmetric and/or a relatively more energetic explosion), or 
that the Cas A yield is in fact "normal", since it is in better agreement with the solar 
$^{44}$Ca/$^{56}$Fe ratio~\cite{renaud06}. 

There are many uncertainties connected to the yield estimate, the most severe ones
being due to astrophysical issues. However, there is also significant uncertainties 
related to the nuclear physics input, in particular the nuclear level density (NLD), 
the radiative strength function (RSF) and the $\alpha$-particle optical model potential. 
All these quantities are entering the calculation of the astrophysical reaction rates.
The NLD is defined as the number of nuclear energy levels
per energy unit at a specific excitation energy, while the RSF gives a measure of the average 
(reduced) transition probability for a given $\gamma$-ray energy. Both quantities are related to the 
average decay probability for an ensemble of levels, and are indispensable in a variety of applications
(reaction rate calculations relevant for astrophysics or transmutation of nuclear waste) as well as for
studying various nuclear properties in the quasicontinuum region.

In this work, we have extracted primary $\gamma$-ray spectra from the decay cascades in $^{44}$Ti
populated via the two-neutron pick-up reaction $^{46}$Ti($p,t\gamma$)$^{44}$Ti.
These spectra are measured for a wide range of initial excitation energies below the neutron threshold,
and they put
a constraint on the functional form of the NLD and RSF. 
%Further, we have compared models
%for the NLD and RSF with our experimental data, and applied the models in calculations of the $\alpha$-capture
%cross section and reaction rate. Further, we have compared the calculations with cross-section data from 
%Nassar \textit{et al.}~\cite{Nassar06} and measured reaction rates 
%from Vockenhuber \textit{et al.}~\cite{vockenhuber07}. 

In Sec.~\ref{sec:exp} we will describe the experimental details, give an overview of the analysis, 
and present the obtained primary $\gamma$-ray spectra.
Different normalizations of the NLD and RSF are discussed in Sec.~\ref{sec:res}, and the applied models
are tested against the primary $\gamma$-ray spectra in Sec.~\ref{sec:reprod}.
In Sec.~\ref{sec:cross}
various input models are used for estimating the 
$^{40}$Ca($\alpha,\gamma$)$^{44}$Ti cross-section and reaction rates, and the calculated results
are compared to existing data from 
Nassar \textit{et al.}~\cite{Nassar06} and 
Vockenhuber \textit{et al.}~\cite{vockenhuber07}. Finally, a summary 
and concluding remarks are given in 
Sec.~\ref{sec:con}.

\section{Experiment and analysis}
\label{sec:exp}

The experiment was conducted at the Oslo Cyclotron Laboratory (OCL), where the Scanditronix cyclotron 
delivered a 32--MeV proton beam bombarding a self-supporting target of $^{46}$Ti with mass thickness
3.0 mg/cm$^2$. The beam current was $\approx 0.5$ nA and the experiment was run for ten days. Unfortunately, the target was
enriched only to 86.0\% in $^{46}$Ti. The main impurities were $^{48}$Ti (10.6\%), $^{47}$Ti (1.6\%), 
$^{50}$Ti (1.0\%), and $^{49}$Ti (0.8\%). The reaction of interest, $^{46}$Ti($p,t\gamma$)$^{44}$Ti, has a
$Q$-value of $-14.236$ MeV~\cite{qcalc}. Particle-$\gamma$ coincidences from this reaction were measured
with eight collimated Si $\Delta E - E$ particle detectors and the CACTUS multidetector system~\cite{CACTUS}. 

The Si
detectors were placed in a circle in forward direction, $45^\circ$ relative to the beam axis. The front 
($\Delta E$) and end ($E$) detectors had a thickness of $\approx 140$ $\mu$m and $1500$ $\mu$m, respectively. 
The CACTUS array consists of 28 collimated $5" \times 5"$ NaI(Tl) crystals for detecting $\gamma$-rays. 
The total efficiency of CACTUS is $15.2(1)$\% at $E_\gamma = 1332.5$ keV. Also a Ge detector was placed
in the CACTUS frame to monitor the experiment.
The charged ejectiles and the 
$\gamma$-rays were measured in coincidence event-by-event.

\begin{figure}[tb]
 \begin{center}
 \includegraphics[clip,width=\columnwidth]{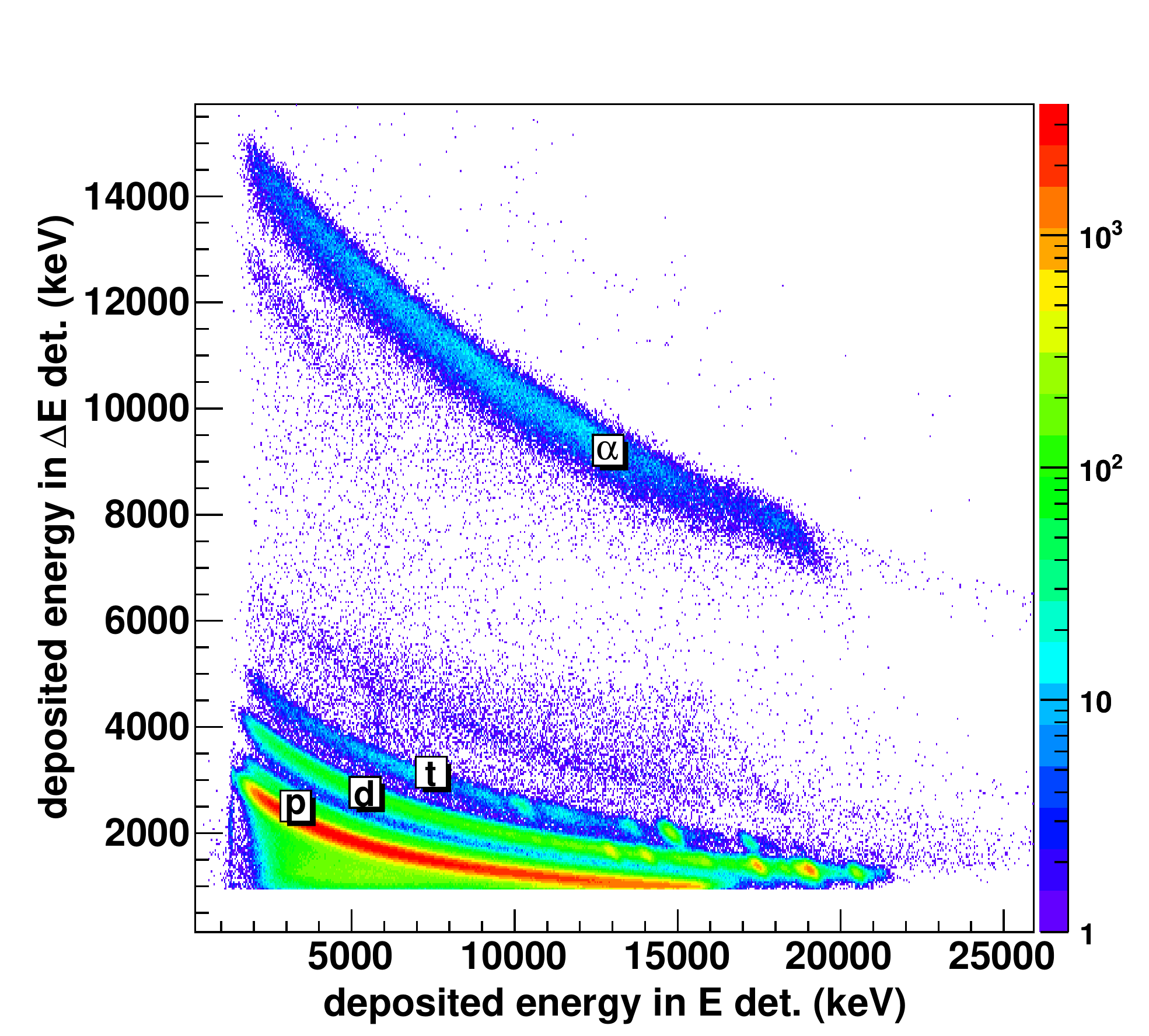}
%\vskip 2cm
 \caption{(Color online). Identification of particle species using the $\Delta E - E$ technique.}
 \label{fig:banana}
 \end{center}
 \end{figure}

To identify the charged ejectiles of the reactions, the well-known $\Delta E -E$ technique is used. 
In Fig.~\ref{fig:banana}, the energy deposited in the $\Delta E$ detector versus the energy deposited
in the $E$ detector is shown for one particle telescope. Each "banana" in the figure corresponds 
to a specific particle species as indicated in the figure.  
%---------------------------------------------------%
%---------------------------------------------------%

\begin{figure}[bt]
 \begin{center}
 \includegraphics[clip,width=\columnwidth]{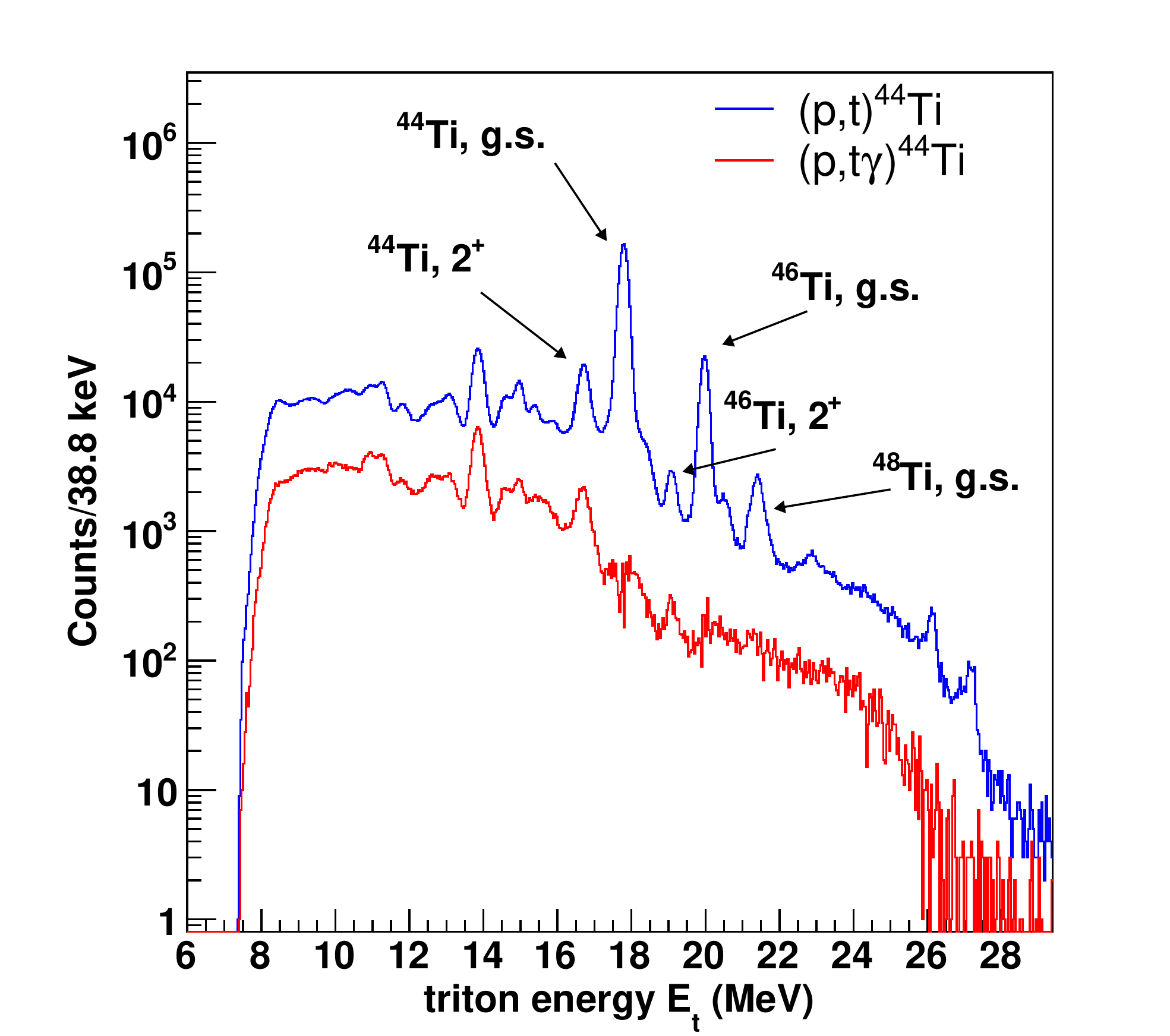}
%\vskip 2cm
 \caption{(Color online). Singles (blue) and coincidence (red) triton spectra. 
 The impurities of $^{48,50}$Ti are easily seen. The resolution is $\approx 270-330$ keV.}
 \label{fig:tritonspectra}
 \end{center}
 \end{figure}

By gating 
on the triton "banana", the events with the reaction $^{46}$Ti($p,t\gamma$)$^{44}$Ti were isolated. 
The $^{46}$Ti($p,d\gamma$)$^{45}$Ti data have been published previously in Ref.~\cite{Naeem_45Ti}. 
The singles and coincidence triton spectra are shown in Fig.~\ref{fig:tritonspectra}. 
We note that the excited 1.904-MeV 0$^{+}$ state ($E_t \approx 16$ MeV)
is rather weakly populated. This is expected on the basis that $\ell=0$ states are in 
general much weaker populated
in the ($p,t$) reaction at $45^\circ$ compared to, e.g., $\ell=2$ and 4, see Ref.~\cite{rapaport}.
The populated spin range is estimated to $J\approx 0-6\hbar$ based on the observed triton peaks
and their coincident $\gamma$-decay cascades, in accordance with the findings in Ref.~\cite{rapaport}.  
%---------------------------------------------------%
%---------------------------------------------------%

Two-neutron pick-up on the $^{48,50}$Ti impurities in the target give rise to peaks at higher
triton energies than the ground state of $^{44}$Ti due to their lower reaction thresholds; the $Q$-values 
are $-12.025$ and $-10.560$ MeV for the $^{44}$Ti($p,t\gamma$)$^{46}$Ti and 
$^{50}$Ti($p,t\gamma$)$^{48}$Ti reactions, respectively. This means that there is a background from such
events in the spectra that cannot be removed (an $\approx 14$\% effect).

Using reaction kinematics and the known $Q$-value for the reaction, the measured triton energy was 
transformed into excitation energy of the residual nucleus. Thus, the $\gamma$-ray spectra are
tagged with a specific initial excitation energy in $^{44}$Ti. Further, the $\gamma$-ray spectra 
were corrected for the known response functions of the CACTUS array following the procedure 
described in Ref.~\cite{gutt6}. The main advantage of this correction method
is that the experimental statistical uncertainties are preserved, without introducing any new, artificial
fluctuations. 

Information on NLD and RSF can be extracted from the distribution of primary $\gamma$-rays, that is, 
the $\gamma$-rays that are emitted first in each decay cascade. To separate these first-generation
$\gamma$ transitions from the second and higher-order generations, an iterative subtraction technique 
is used~\cite{Gut87}. The $\gamma$-ray spectra $f_i$ for excitation-energy bin $i$ obviously contain 
all generations of $\gamma$-rays from all possible cascades decaying from the excited states of this bin. 
The subtraction technique is based on the assumption that the spectra $f_{j<i}$ for all the energy 
bins $E_j < E_i$ contain the same $\gamma$-transitions as $f_i$ \textit{except} the first $\gamma$-rays emitted,
since they will bring the nucleus from the states in bin $i$ to underlying states in the energy bins $j$. 
This is true if the main assumption of this technique holds: that the decay routes are the same whether 
they were initiated directly by the nuclear reaction or by $\gamma$ decay from higher-lying states.

\begin{figure}[bt]
 \begin{center}
 \includegraphics[clip,width=\columnwidth]{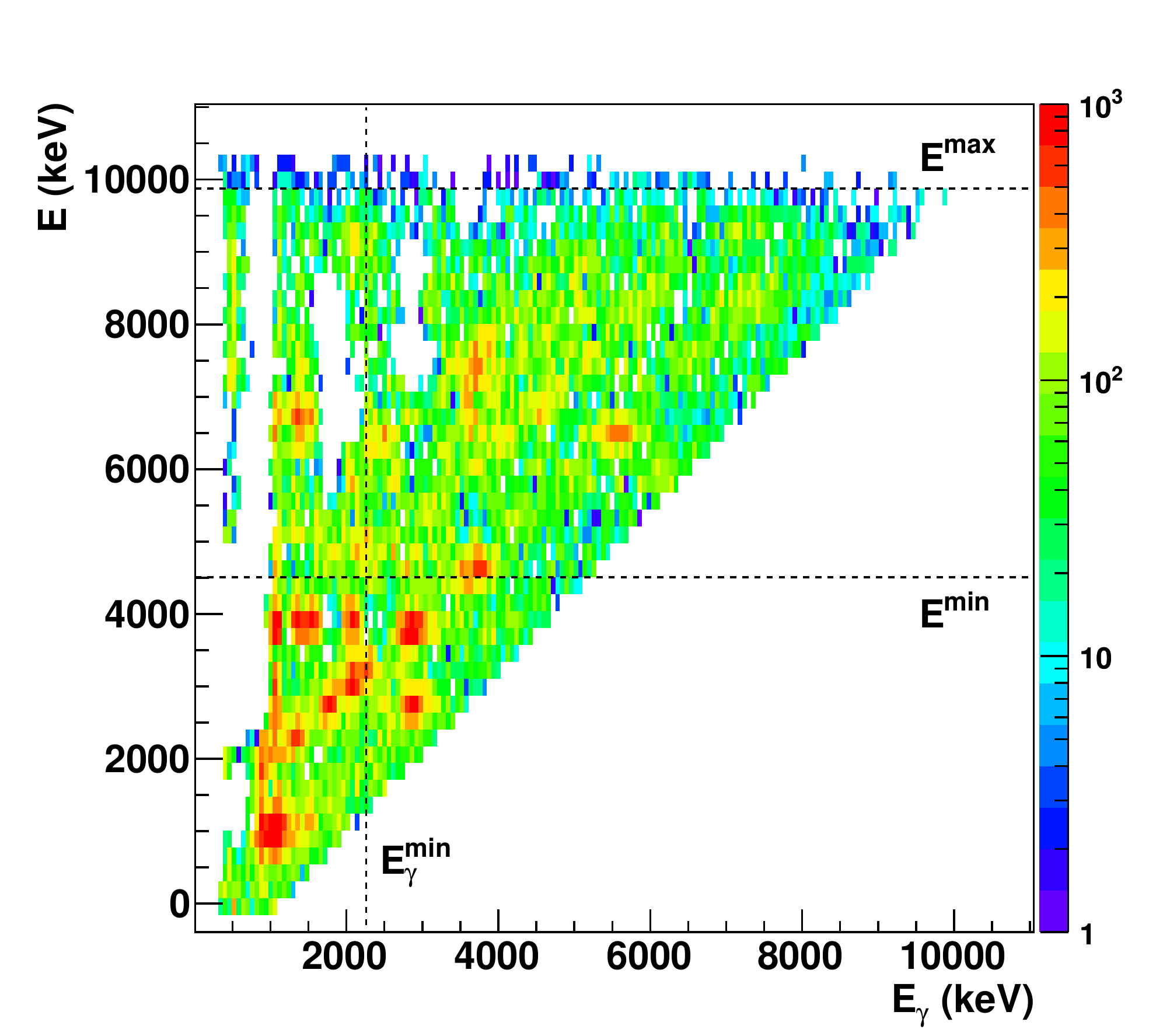}
%\vskip 2cm
 \caption{(Color online). The first-generation spectra for each excitation-energy 
 bin in $^{44}$Ti. The dashed lines are limits set for the further analysis (see text).}
 \label{fig:fgmatrix}
 \end{center}
 \end{figure}

The obtained first-generation matrix $P(E,E_\gamma)$ of $^{44}$Ti is shown in Fig.~\ref{fig:fgmatrix}. 
%---------------------------------------------------%
%---------------------------------------------------%
The diagonal where $E=E_\gamma$ is clearly seen. Here, two peaks are particularly pronounced: 
one from the decay of the 
first excited state at $E = E_\gamma = 1083$ keV, and the other from the decay of the second $2^{+}$
state at $2887$ keV to the ground state. There are also more diagonals visible, e.g., for 
$E_\gamma = E - 1083$ keV where the decay goes directly to the first excited $2^{+}$ state,
and so on. 

\begin{figure}[bt]
 \begin{center}
 \includegraphics[clip,width=\columnwidth]{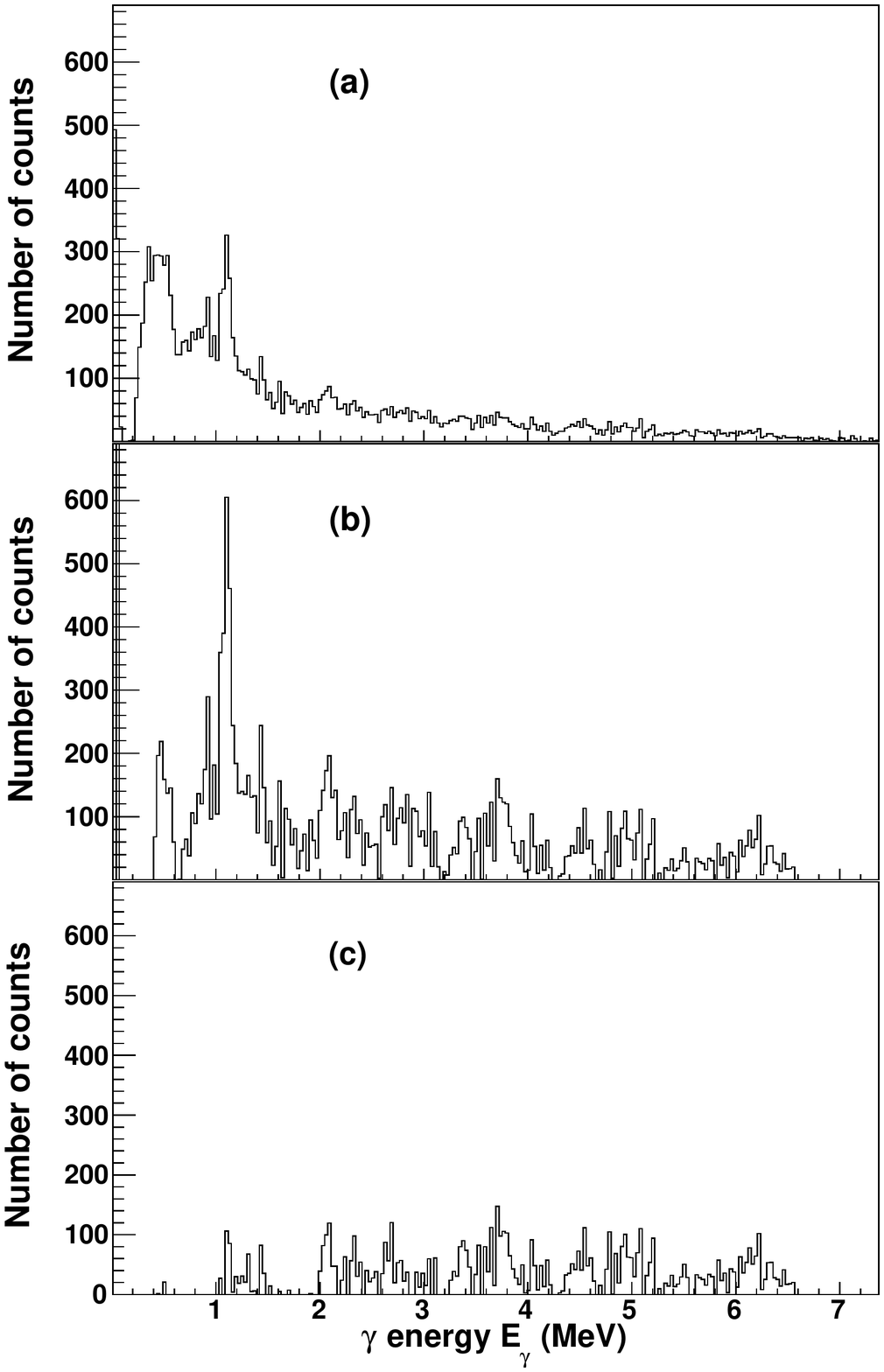}
%\vskip 2cm
 \caption{Original (a), unfolded (b) and first-generation (c) 
 $\gamma$-ray spectrum of $^{44}$Ti for excitation energy $E= 6.1$ MeV.}
 \label{fig:gammaspectra}
 \end{center}
 \end{figure}

Before extracting the NLD and RSF from the $P$ matrix, we have set a lower limit for the $\gamma$-ray
energies ($E_\gamma^{\mathrm{min}}$), and a lower and upper limit for the excitation energy 
($E_{\mathrm{min}}$, $E_{\mathrm{max}})$. The limits on the excitation-energy side are put to ensure
that the spectra are dominated by decay from compound states ($E_{\mathrm{min}}$), and that the 
statistics is not too low
($E_{\mathrm{max}}$). On the $\gamma$-ray energy side the limit is set to exclude possible left-overs
of higher-generation decay, that might not be correctly subtracted in the first-generation
method (see Refs.~\cite{systematic,Schiller00} and references therein for more details). 
In Fig.~\ref{fig:gammaspectra}, the original, unfolded, and first-generation $\gamma$-ray spectrum 
of $^{44}$Ti for excitation energy $E= 6.1$ MeV are displayed. 
%---------------------------------------------------%
%---------------------------------------------------%

With the first-generation matrix properly prepared, an iterative procedure is applied to extract
NLD and RSF from the primary $\gamma$-ray spectra. The method is based on the assumption that 
the reaction leaves the product nucleus in a compound state, which then subsequently decays in a manner
that is independent on the way it was formed, i.e. a statistical decay process~\cite{BM}. 
This is a reasonable 
assumption for states in the quasicontinuum, where the typical lifetime of the states is of the 
order of $10^{-15}$ $s$, whereas the reaction time is $\approx 10^{-18}$ $s$. In addition, the 
configuration mixing of the levels is expected to be significant if the level spacing is comparable to 
the residual interaction~\cite{BM}. This is normally fulfilled in the region of high level density.

If compound states are indeed populated, 
the $\gamma$ decay from these states should be independent of the reaction used to reach them. In previous 
works (e.g. Ref.~\cite{Mo_nld}) it has been demonstrated that the direct reactions 
($^3$He,$^3$He$^\prime$) and ($^3$He,$\alpha$) into the same final nucleus 
do produce very similar decay cascades. Also, the extracted 
NLD and RSF was found to be equal within the expected fluctuations\footnote{Porter-Thomas fluctuations 
and statistical uncertainties must be taken into account.} This indicates that even though a 
direct reaction such as ($p,t$) is used, compound states are likely to be populated at sufficiently 
high excitation energy (as mentioned already for the $E_{\mathrm{min}}$ limit set in the 
first-generation matrix $P(E,E_\gamma)$).

The ansatz for the iterative method is~\cite{Schiller00}:
\begin{equation}
P(E, E_{\gamma}) \propto  \rho (E_{\mathrm{f}}) {\mathcal{T}}  (E_{\gamma}),
\label{eq:brink}
\end{equation}
meaning that the first-generation matrix $P(E,E_\gamma)$ is assumed to be separable into two vectors that give
directly the functional form of the level density at the final excitation energy 
$E_{\mathrm{f}} = E - E_\gamma$, and the $\gamma$-ray transmission coefficient $\cal{T}$ for a given 
$E_\gamma$. 
This is done by minimizing 
\begin{equation}
\chi^{2} = \frac{1}{N_{\mathrm{free}}}\sum_{E=E^{\mathrm{min}}}^{E^{\mathrm{max}}}
	\sum_{E_{\gamma}=E_{\gamma}^{\mathrm{min}}}^{E} 
	\left( \frac{P_{\mathrm{th}}(E, E_{\gamma}) - P(E, E_{\gamma})}{\Delta P(E, E_{\gamma})} \right)^{2},
\label{eq:chisquared}
\end{equation}
where $N_{\mathrm{free}}$ is the number of degrees of freedom, $\Delta P(E, E_{\gamma})$ is the 
uncertainty in the experimental first-generation $\gamma$-ray matrix $P(E,E_\gamma)$, and the theoretical
first-generation matrix is given by 
\begin{equation}
P_{\mathrm{th}}(E, E_{\gamma}) = \frac {\rho (E -E_{\gamma}) {\mathcal{T}}  
(E_{\gamma})}{\sum_{E_{\gamma}=E_{\gamma}^{\mathrm{min}}}^{E} \rho (E -E_{\gamma}) {\mathcal{T}}  (E_{\gamma})}.
\label{eq:theory}
\end{equation}
Every point of the $\rho$ and ${\mathcal{T}}$ functions is assumed to be an independent 
variable, so that the reduced $\chi^{2}$ of Eq.~(\ref{eq:chisquared}) is minimized for every argument 
$E-E_{\gamma}$ and $E_{\gamma}$. 
Note that $\cal{T}$ is independent of excitation energy according to the Brink 
hypothesis~\cite{brink}. If we had an excitation-energy dependent $\gamma$-ray transmission 
coefficient, ${\cal T} = {\cal T}(E,E_{\gamma})$, it would in principle be impossible to 
disentangle the level density and the $\gamma$-ray transmission coefficient. 
%Obviously, at low excitation 
%energy, the $\gamma$ decay is highly dependent on the initial and final state;
%therefore, we have excluded the data below $E^{\mathrm{min}}$ in this procedure. 

It is well known that the Brink hypothesis is violated when high temperatures and/or spins 
are involved in the nuclear reactions, as shown for giant dipole resonance (GDR) excitations in 
Ref.~\cite{Andreas&Thoennessen} and references therein. However, since both the temperature reached 
($T \propto \sqrt{E_{\mathrm{f}}}$) and the spins populated ($J \sim 0 - 6 \hbar$) are rather low for 
the experiment in this work, these dependencies are assumed to be of minor importance in the 
excitation-energy region of interest here.

\begin{figure*}[htb]
 \begin{center}
 \includegraphics[clip,totalheight=11cm]{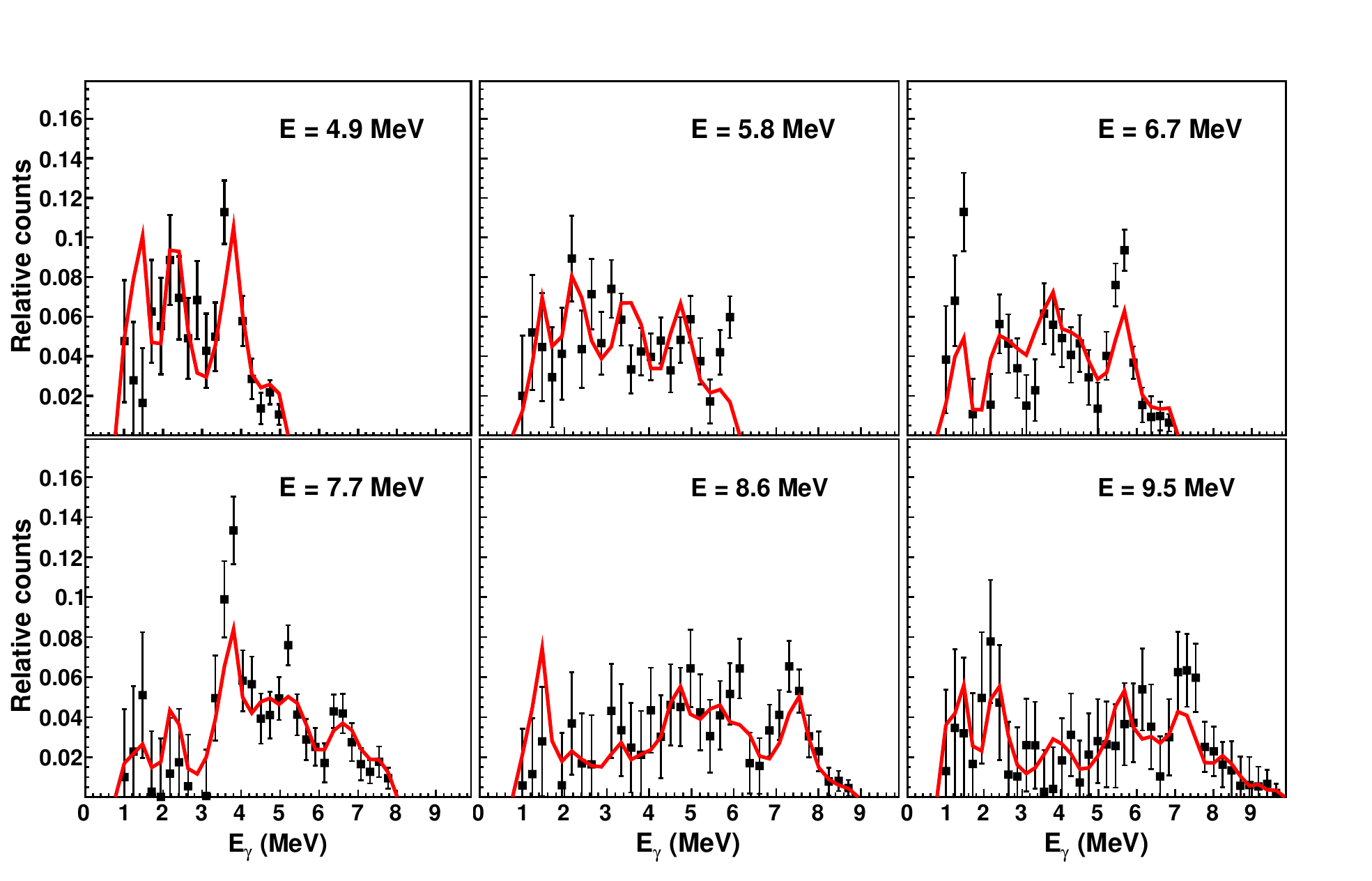}
%\vskip 2cm
 \caption{(Color online). Comparison of experimental primary $\gamma$-ray spectra (filled squares) and the
 ones obtained from multiplying the extracted $\rho$ 
 and $\cal{T}$ functions (red line) for several excitation energies. The experimental 
 and calculated spectra are shown for excitation-energy bins of 467 keV.}
 \label{fig:work}
 \end{center}
 \end{figure*}

To inspect how well the iterative procedure works, we have compared the experimental first-generation 
spectra for several excitation energies with the ones obtained by multiplying the extracted $\rho$ 
and $\cal{T}$ functions. The result for a selection of primary $\gamma$-ray spectra is shown in Fig.~\ref{fig:work}. 
%---------------------------------------------------%
%---------------------------------------------------%
%All experimental first-generation spectra considered are shown in the left part of Fig.~\ref{fig:fgteo},
%while the right part shows the calculated spectra using the extracted $\rho$ and $\cal{T}$ functions.
As can be seen, 
%from these two figures, 
the agreement between the calculated 
and the experimental first-generation spectra are in 
general quite good, although there are local variations where the calculated spectra are
not within the error bars of the experimental ones. These variations could well be due to 
large Porter-Thomas fluctuations~\cite{PT}, as there are relatively few levels in this nucleus. 
%However, the global $\chi^2$ result for all the spectra is acceptable ($\chi^2 = 1.8$).  
%---------------------------------------------------%
% \begin{figure*}[hbt]
% \begin{center}
% \includegraphics[clip,totalheight=7cm]{fgrsg.pdf}
% \includegraphics[clip,totalheight=7cm]{fgteorsg.pdf}
%\vskip 2cm
% \caption{(Color online). Left: experimental first-generation spectra. Right: calculated spectra.}
% \label{fig:fgteo}
% \end{center}
% \end{figure*}
%---------------------------------------------------%

The iterative procedure to obtain the level density and the $\gamma$-ray transmission 
coefficient uniquely determines the functional form of $\rho$ and $\cal{T}$; however,
identical fits to the experimental
data is achieved with the transformations~\cite{Schiller00} 
\begin{eqnarray}
\tilde{\rho}(E-E_\gamma)&=&A\exp[\alpha(E-E_\gamma)]\,\rho(E-E_\gamma),
\label{eq:array1}\\
\tilde{{\mathcal{T}}}(E_\gamma)&=&B\exp(\alpha E_\gamma){\mathcal{T}} (E_\gamma).
\label{eq:array2}
\end{eqnarray}
Thus, to obtain the absolute normalization of 
the level density and $\gamma$-ray transmission 
coefficient, 
the transformation parameters $A$, $\alpha$, and $B$ must be determined independently.

\section{Normalizations and models for\newline level density and radiative strength function}
\label{sec:res}
%The shape of the first-generation spectra restrict the possible combinations of input level
%density and RSF in, e.g., cross section calculations. We will in the following normalize 
%the extracted level density and RSF in a coherent way using standard input models for these
%quantities.
\subsection{Level density}
\label{subsec:nld}
Usually, the level density $\rho(E_f)$ is normalized to known, discrete levels at low excitation energy, and
to the total level density at the neutron separation energy $\rho(S_n)$, which is deduced from neutron 
resonance spacings $D_0$ and/or $D_1$ (see, e.g., Ref.~\cite{Pb}).
%, and assuming a spin distribution according 
%to Ref.~\cite{GC}: 
%\begin{equation}
%g(E,I) \simeq \frac{2I+1}{2\sigma^2}\exp\left[-(I+1/2)^2/2\sigma^2\right].
%\label{eq:spindist}
%\end{equation}
However, for the $^{44}$Ti case, no neutron (or proton) resonance data are known since the target nuclei
for the neutron and proton capture reactions are unstable. We are therefore left with only two types of 
experimental constraints, namely the known levels at low excitation energy and the observed 
first-generation spectra (which also depend on the $\gamma$-ray strength). 

\begin{figure}[ht]
 \begin{center}
 \includegraphics[clip,width=\columnwidth]{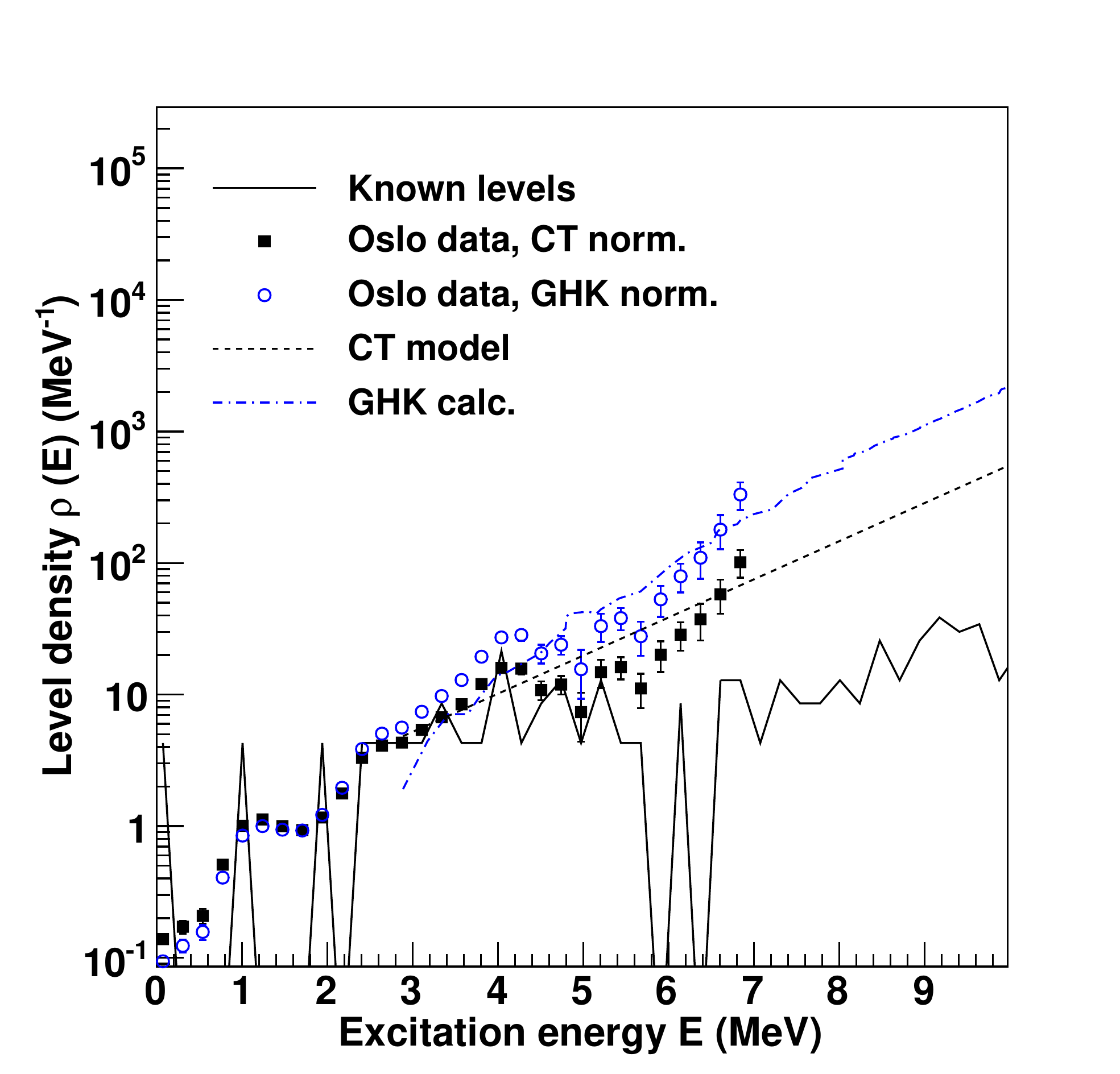}
%\vskip 2cm
 \caption{(Color online) Experimental data points (filled squares) normalized to the CT model (dashed line)
 and data points (blue, open circles) normalized to the calculation of GHK (dashed-dotted line). 
 The black, solid line represents the known levels taken from Ref.~\cite{ENSDF}. }
 \label{fig:nldnorm}
 \end{center}
 \end{figure}

We have chosen two different approaches to normalize our data, and, out of those, 
to estimate the sensitivity on our results to this normalization procedure: 
(\textit{i}) apply theoretical level densities based on microscopic calculations,
and (\textit{ii}) use a standard closed-form formula with global parameters.
For the first approach, we have used recent calculations of Goriely, Hilaire and Koning~\cite{go08} (hereafter labeled GHK). 
For the second approach, we have applied the constant-temperature (CT) formula~\cite{GC}: 
\begin{equation}
\rho_{\mathrm{CT}}(E) = \frac{1}{T}\exp{\left( \frac{E-E_0}{T} \right)},
\end{equation}
where the nuclear temperature $T=1.50$ MeV and the energy shift $E_0=-0.08$ MeV are taken from the global parameterization 
of Refs.~\cite{koning08,TALYS}. The level-density data normalized to these 
two approaches are shown in Fig.~\ref{fig:nldnorm}. 
%---------------------------------------------------%
%---------------------------------------------------%

%************************************************************************************%
%\begin{table}[htb]
%\caption{Parameters used for the level-density models.} 
%\begin{tabular}{llccc}
%\hline
%\hline
%Model     & $E_{0},\,\Delta^{\mathrm{BSFG}}$ & $T$   & $a$          & $\sigma$  \\
%		  & (MeV)   & (MeV) & (MeV$^{-1}$) &           \\
%\hline
%CT        & -0.080  & 1.50  &  $-$         & $-$      \\
%BSFG      & -0.670  & $-$   & 5.47         & 2.93      \\
%Comb.     &  $-$    & $-$   &    $-$       & $-$       \\
%\hline
%\hline
%\end{tabular}
%\\
%\label{tab:nldpar}
%\end{table}
% & $D_0$ & $\rho(S_p)$ 
% & (keV) & (MeV$^{-1}$)
% &  25   &   226       
% &  14   &   390       
% &  3.7  &   914       
%************************************************************************************%

%Note that the CT model gives very similar results to the global parameterization of 
%von Egidy and Bucurescu~\cite{egidy2} for the Back-Shifted Fermi Gas model (BSFG), which is applied
%for $^{45}$Ti in~\cite{Naeem_45Ti}. 
We observe that our data follow closely the known, discrete levels~\cite{ENSDF} 
at low excitation energy, and especially in the region $2.4 < E < 5.2$ MeV. This is gratifying, as it 
implies that the Oslo method does indeed give reasonable results for the level density. We also see that
there is a decrease in the level density for $ 4.0 < E < 5.2$ MeV and an abrupt increase  for $ 5.8 < E < 6.2$ MeV.
These structures are not well described by any of the models used for normalization. They might be due to
shell effects and/or $\alpha$-clustering effects (since $^{44}$Ti is an $N=Z$ nucleus). An increase in
the level density could also
indicate the breaking of a nucleon Cooper pair and/or the crossing of a shell gap. 
%By looking at the 
%estimated pair-gap parameters evaluated from even-odd mass differences \cite{Wapstra} according to Ref.~\cite{BM}, 
%one finds that $2\Delta_n = 5.40$ MeV and $2\Delta_p = 5.38$ MeV.  

We note that both the ground state and the 0$^{+}$ state at $E=1.9$ MeV are not very 
pronounced in Fig.~\ref{fig:nldnorm}.
Also, we see that
there is less direct decay to the ground state and the excited 0$^{+}$ state compared to the 2$^{+}$ state at
1083 keV and the 4$^{+}$ state at 2454 keV, especially at higher excitation energies (see Fig.~\ref{fig:fgmatrix}). 
This could imply that there are relatively few
spin-1 states populated in the ($p,t$) reaction (assuming that dipole radiation is dominant in this region). 
This is not surprising because the spin distribution is expected to have a maximum for $J=3-4$. 

We see that the resolution on the level density is rather poor at low excitation energies, roughly 600 keV for the 
first excited 2$^{+}$ state. This may be explained by the fact
that the level density in this region is mainly determined by decay involving high-energy $\gamma$ rays, which have 
a resolution of up to $\approx 300$ keV (similar to the triton-energy resolution). As a consequence, the resolution 
of the level density gets better and better for increasing excitation energy. In addition, the rather large amount of other Ti 
isotopes in the target may give a smoothing effect on the extracted quantities (both the NLD and RSF).

\subsection{Radiative strength function}
\label{sec:rad}
\begin{figure}[bt]
 \begin{center}
 \includegraphics[clip,width=\columnwidth]{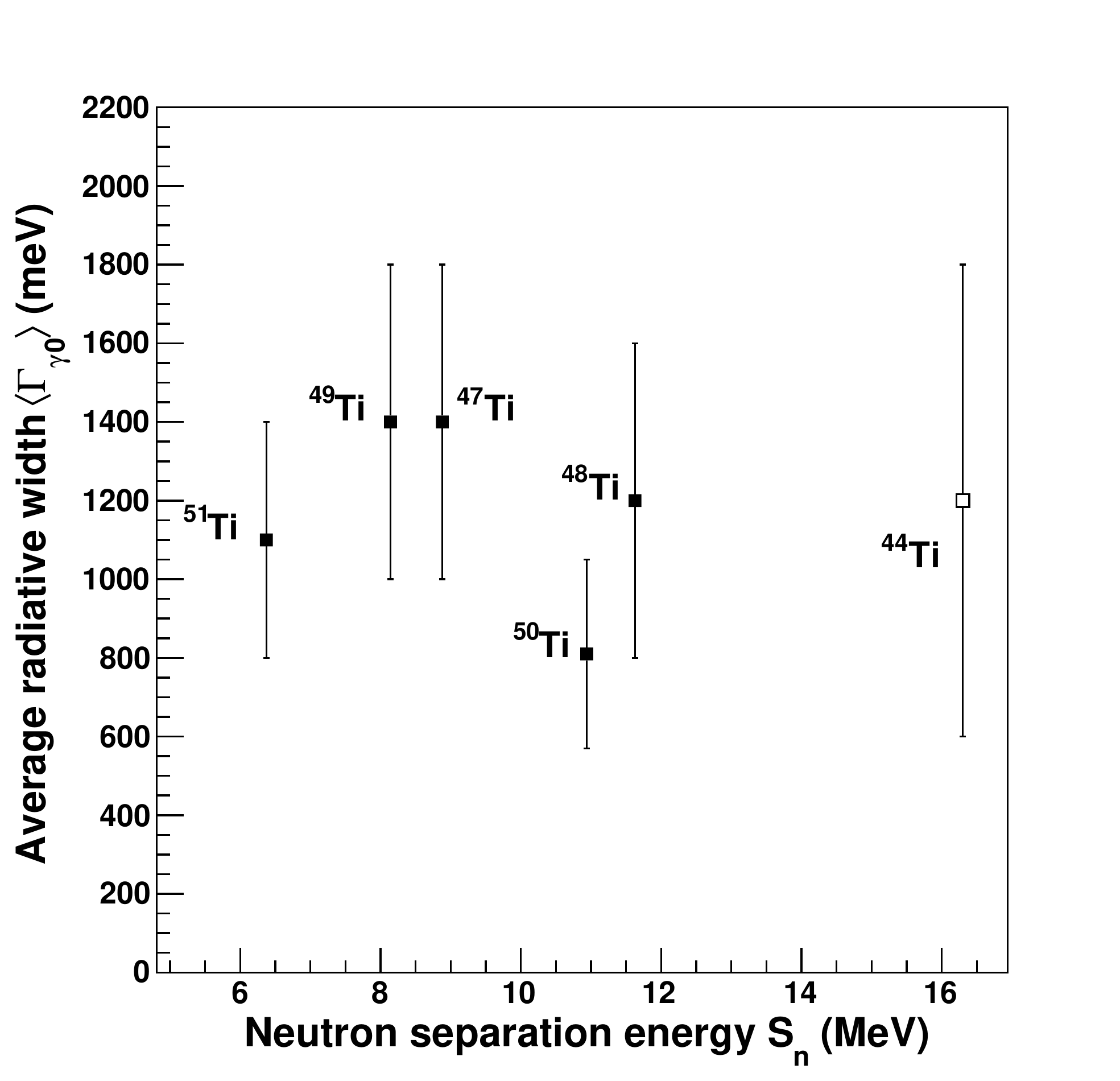}
%\vskip 2cm
 \caption{Experimental average radiative width (filled squares) for $^{47-51}$Ti as a function of neutron separation
	energy, and estimated width for $^{44}$Ti (open square). }
 \label{fig:gammasyst}
 \end{center}
 \end{figure}

\begin{figure}[bt]
 \begin{center}
 \includegraphics[clip,width=\columnwidth]{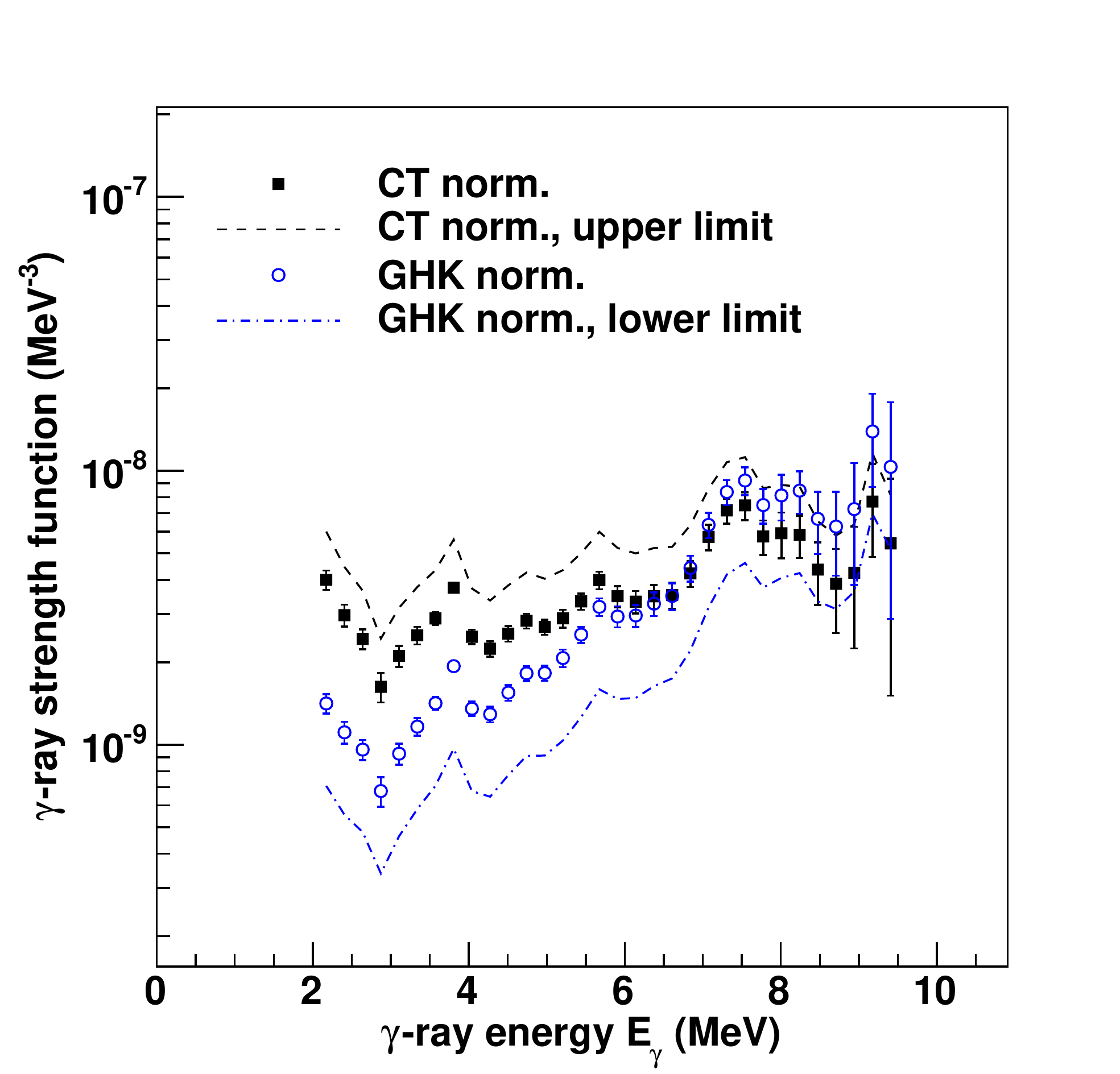}
%\vskip 2cm
 \caption{(Color online). Normalized radiative strength functions for the two normalizations
	of the level density. The dashed line is the upper limit for the CT normalization,
	while the dashed-dotted line is the lower limit for the GHK normalization.}
 \label{fig:rsfs}
 \end{center}
 \end{figure}

Because there are no experimental data on neither the level spacing $D$ nor the total, average  
radiative width for s-wave resonances $\left<\Gamma_{\gamma0}\right>$ for $^{44}$Ti, we must find the absolute normalization
parameter $B$ of the RSF by other means. In order to have a rough approximation of the absolute strength, 
we have looked at experimental values of  $\left<\Gamma_{\gamma0}\right>$ for other Ti 
isotopes found in Ref.~\cite{RIPL2}, see Fig.~\ref{fig:gammasyst}. From these values, 
the educated guess of $\left<\Gamma_{\gamma0}\right> = 1200(600)$ meV seems to be reasonable for $^{44}$Ti. 
The RSFs of $^{44}$Ti for the two choices of level-density normalization are shown in Fig.~\ref{fig:rsfs}. Also the upper 
and lower limits are indicated, corresponding to $\left<\Gamma_{\gamma0}\right> = 1800$ meV for the CT normalization
and $\left<\Gamma_{\gamma0}\right> = 600$ meV for the GHK normalization, respectively. 
%---------------------------------------------------%
%---------------------------------------------------%
%---------------------------------------------------%
%---------------------------------------------------%

From Fig.~\ref{fig:rsfs}, it is seen that for $E_\gamma > 4$ MeV, the RSF of $^{44}$Ti seems to reach a relatively smooth
behavior with increasing strength as a function of $E_\gamma$. Naturally, as seen from Eq.~\ref{eq:array2}, 
the slope varies depending on the 
normalization chosen for the level density. For energies below $E_\gamma \approx 4$ MeV, we observe 
on average a slight increase in strength
for decreasing $\gamma$-ray energy. However, we see that the data in this region display quite large variations,
which could be due to Porter-Thomas fluctuations~\cite{PT}.

A low-energy increase in the RSF data has been seen previously in several light and medium-mass nuclei with 
the Oslo method~\cite{Fe_Alex,Fe_Emel,Mo_RSF,V,Sc}, with the two-step cascade method following neutron 
capture~\cite{Fe_Alex}, and recently also in proton capture~\cite{Ni_Alex}. However, for $^{44}$Ti, the increase at low energies is not 
as strong as in, e.g., $^{56,57}$Fe~\cite{Fe_Alex}. 

As of today, it is not clear whether the low-energy increase is due to some sort of collective decay mode(s) or if it is
an effect of other structural effects in these nuclei. An analysis of simulated data using the DICEBOX code~\cite{DICEBOX}
has demonstrated that for light nuclei, 
the spin distribution
of the initial populated levels may have a considerable influence on the possible decay paths from these levels~\cite{systematic}. 
This seems to be due to a combination of several factors: (\textit{i}) the low level density in light nuclei at low excitation energy, 
(\textit{ii}) restrictions on the possible populated spins of the initial levels, 
(\textit{iii}) the dominance of dipole radiation from highly excited levels, and (\textit{iv}) a rather large asymmetric
parity distribution up to rather high excitation energies. As is shown in Ref.~\cite{systematic}, these factors may lead to 
a significant increase in the extracted RSF for low $\gamma$-ray energies compared to the input RSF 
function used to generate the $\gamma$-ray spectra. It is not unlikely that a similar effect may be present in $^{44}$Ti
as well.

%However, if one insists on using the ansatz in Eq.~(\ref{eq:brink})
%to obtain a reasonable reproduction of the simulated primary $\gamma$ spectra, it is necessary to use 
%the extracted RSF, which displays this low-energy enhancement. 
%This means that the influence of the four factors mentioned above 
%on the decay probability is effectively taken care of by including this enhancement.

\subsection{Models for the radiative strength function}
\label{sec:mod}
In general, %it is believed that
the $\gamma$ decay in quasicontinuum is expected to be dominated by electric dipole radiations. Also, there is experimental
evidence that a giant magnetic dipole resonance (also known as the magnetic spin-flip resonance) is present in 
several light and medium-mass nuclei (see, e.g., Ref.~\cite{Djalali}). 
%We will in the following describe 
%some models that, with certain parameterizations, are able to describe our experimental data reasonably well. 

We have applied the commonly used Generalized Lorentzian (GLO)  
expression~\cite{ko87,ko90} for the $E1$ strength.
This model is % based on the theory of
%Fermi liquids and accounts for microscopic properties of the GDR. It is designed to reproduce $\gamma$-ray intensities from average 
%resonance capture reactions as well as photonuclear cross-section data, and is 
given by~\cite{ko90}
\begin{align}
& f_{\rm GLO}(E_{\gamma},T_f) = \frac{1}{3\pi^2\hbar^2c^2}\sigma_{E1}\Gamma_{E1} \times \\ \nonumber
& \left[\frac{ E_{\gamma} \Gamma(E_{\gamma},T_f)}{(E_\gamma^2-E_{E1}^2)^2 + E_{\gamma}^2 
	\Gamma (E_{\gamma},T_f)^2} + \;0.7\frac{\Gamma(E_{\gamma}=0,T_f)}{E_{E1}^3} \right],
\label{eq:GLO}
\end{align} 
where the Lorentzian parameters $\Gamma_{E1}$, $E_{E1}$ and $\sigma_{E1}$ correspond to the width, centroid energy, and peak cross section of the giant electric dipole resonance (GDR). We have made use of the parameterization 
of RIPL-2~\cite{RIPL2} to estimate the GDR parameters as these are unknown experimentally. 
Due to the dynamic ground-state deformation of $^{44}$Ti 
($\beta_2 = 0.27$~\cite{RIPL2}), the GDR is assumed to be split in two components and thus two sets of
GDR parameters are applied (see Tab.~\ref{tab:rsfpar}). 
In addition, we have assumed a constant temperature $T_f$ of the 
final states in accordance with the Brink hypothesis~\cite{brink}. 
%It should be noted that the GLO model
%resembles the Kadmenski{\u{\i}}, Markushev and Furman (KMF) model~\cite{ka83}, which is based on Fermi-liquid theory, 
%in the energy region studied here. 

In order to account for the small increase in
strength at low energies, we have %assumed that it is of $E1$ character and have 
added a Lorentzian resonance of the form 
\begin{equation}
f_{\mathrm{up}}(E_\gamma)=\frac{1}{3\pi^2\hbar^2c^2} 
	\frac{\sigma_{\mathrm{up}}E_\gamma\Gamma_{\mathrm{up}}^2} {(E_\gamma^2-E_{\mathrm{up}}^2)^2+E_\gamma^2\Gamma_{\mathrm{up}}^2},
\label{eq:upb}
\end{equation}
with centroid energy $E_{\mathrm{up}}=1.9$~MeV, width $\Gamma_{\mathrm{up}}=1.3$~MeV, and a peak cross section 
$\sigma_{\mathrm{up}}$ that will vary according to which NLD model that has been used for normalization. Such a shape of the 
low-energy increase seems to be in accordance with data and model descriptions in Refs.~\cite{milan,steve} for the Mo nuclei, 
and was used also in Ref.~\cite{upbend_cross}. The 
constant temperature $T_f$ is slightly varied for the two models to give the best fit to the data.
All parameters used are given in Tab.~\ref{tab:rsfpar}. 

The lower and upper limit of the absolute normalization will necessarily give slightly different parameters to get 
the best fit to the data. These parameters are also given in Tab.~\ref{tab:rsfpar}.

Another frequently used model for $E1$ strength is the standard Lorentzian (the Brink-Axel model, see Ref.~\cite{RIPL2}
and references therein). The expression reads
\begin{equation}
f_{\mathrm{SLO}}(E_\gamma)=\frac{1}{3\pi^2\hbar^2c^2} 
	\frac{\sigma_{E1}E_\gamma\Gamma_{E1}^2} {(E_\gamma^2-E_{E1}^2)^2+E_\gamma^2\Gamma_{E1}^2}.
\label{eq:stLor}
\end{equation}
This model is independent of excitation energy, in accordance with our assumption 
behind Eq.~(\ref{eq:brink}). However, as described in Ref.~\cite{RIPL2}, this model is known to
generally overestimate the value of $\left<\Gamma_{\gamma0}\right>$ and neutron-capture cross sections. 

\begin{figure}[bt]
 \begin{center}
 \includegraphics[clip,width=\columnwidth]{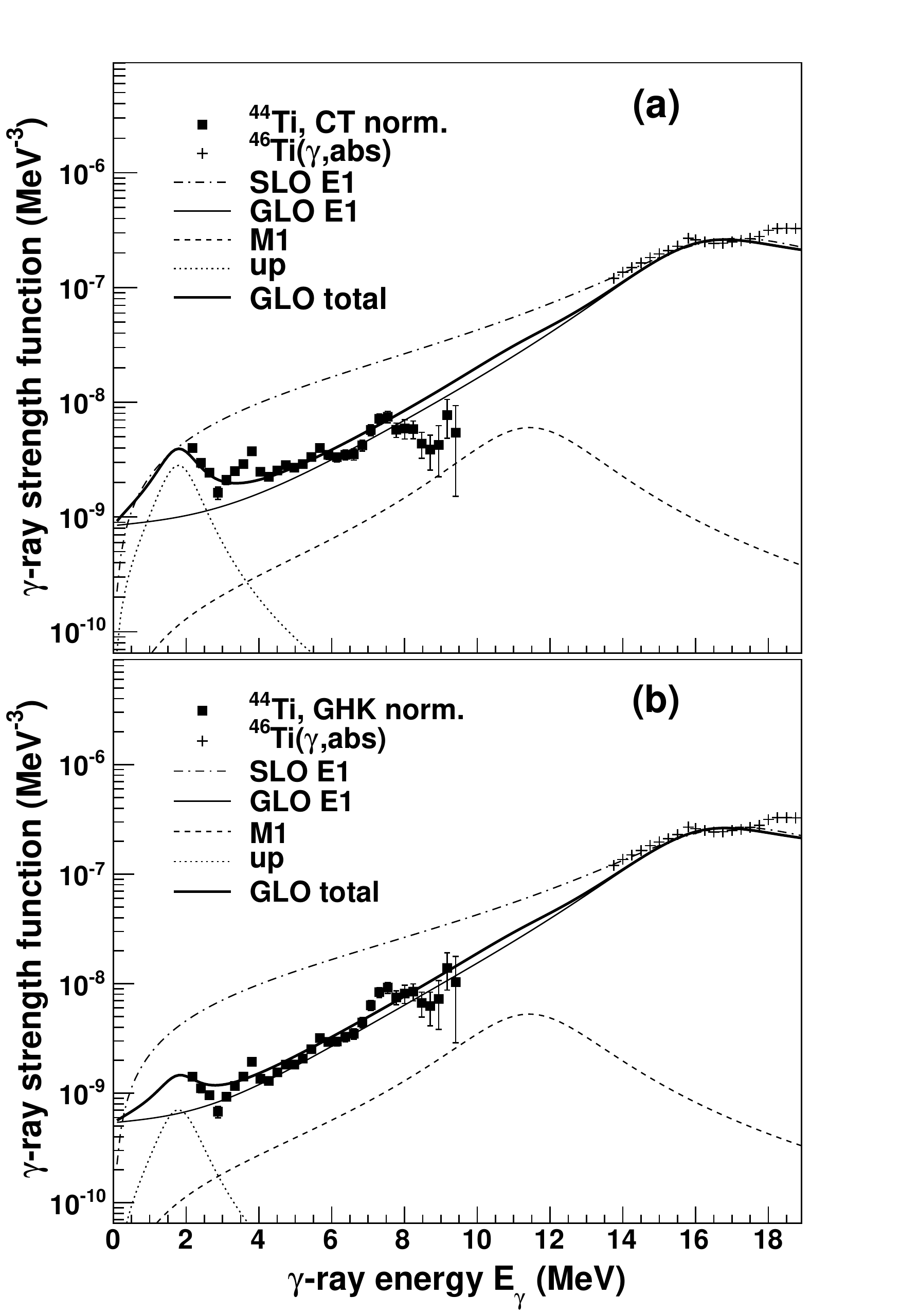}
%\vskip 2cm
 \caption{Radiative strength functions 
	and fitted models for (a) the CT and (b) the GHK normalization.
	Note that the SLO model has not been fitted to the data.}
 \label{fig:rsfmodels}
 \end{center}
 \end{figure}

For the magnetic transitions, a standard Lorentzian as recommended by the RIPL-2 library \cite{RIPL2} 
and shown in Fig.~\ref{fig:rsfmodels} is adopted (see Table~\ref{tab:rsfpar} for the corresponding parameters). 
%STEPH: I would skip this discussion on the M1, it is relatively standard and complicates just the manuscript !
%we have applied a standard Lorentzian of the form
%\begin{equation}
%f_{\mathrm{M1}}(E_\gamma)=\frac{1}{3\pi^2\hbar^2c^2} 
%	\frac{\sigma_{M1}E_\gamma\Gamma_{M1}^2} {(E_\gamma^2-E_{M1}^2)^2+E_\gamma^2\Gamma_{M1}^2}.
%\label{eq:magnetic}
%\end{equation}
%with Lorentzian parameters $\sigma_{M1}$, $\Gamma_{M1}$, and $E_{M1}$, which correspond to the peak cross section, 
%the full-width half maximum (FWHM), and the centroid of the resonance. Using systematics according to Ref.~\cite{RIPL2}
%gave in this case a very large $M1$ component. Because the spin-flip $M1$ resonance is poorly known compared 
%to the GDR and the $E2$ isoscalar resonance, we have taken the liberty to reduce the peak cross section $\sigma_{M1}$,
%see Tab.~\ref{tab:rsfpar} for the applied parameters. Thus, we get 
%an $M1$ contribution of about 15\% of the $E1$ strength at $E_\gamma = 11.6$ MeV.
%************************************************************************************%
\begin{table*}[htb]
\caption{Parameters used for the RSF models.} 
\begin{tabular}{lccccccccccccc}
\hline
\hline
Model    & $E_{E1,1}$ & $\sigma_{E1,1}$ & $\Gamma_{E1,1}$ & $E_{E1,2}$ & $\sigma_{E1,2}$ & $\Gamma_{E1,2}$ & $T_f$ & $E_{M1}$ & $\sigma_{M1}$ & $\Gamma_{M1}$  & $E_{\mathrm{up}}$ & $\sigma_{\mathrm{up}}$ & $\Gamma_{\mathrm{up}}$ \\
		 & (MeV)  	& (mb)          &  (MeV)        & (MeV)  	& (mb)          &  (MeV) & (MeV)      & (MeV)    & (mb)          & (MeV)          & (MeV)             & (mb)                   & (MeV)                  \\
\hline
CT+GLO   & 17.23     & 43.16          & 5.98          & 21.58     & 21.54          & 9.19 & 0.50        & 11.6     & 0.8          & 4.0            & 1.9               & 0.060                   & 1.3                    \\
CT+GLO, upper limit   & 17.23     & 43.16          & 5.98          & 21.58     & 21.54          & 9.19 & 0.80        & 11.6     & 0.9          & 4.0            & 1.9               & 0.070                   & 1.3                    \\
%BSFG+GLO & 17.23     & 43.16          & 5.98          & 21.58     & 21.54          & 9.19 & 1.1        & 11.6     & 1.0          & 4.0            & 1.3               & 0.40                   & 1.1                    \\
GHK+GLO & 17.23     & 43.16          & 5.98          & 21.58     & 21.54          & 9.19 & 0.40        & 11.6     & 0.7          & 4.0            & 1.9               & 0.015                   & 1.3                    \\
GHK+GLO, lower limit & 17.23     & 43.16          & 5.98          & 21.58     & 21.54          & 9.19 & 0.15        & 11.6     & 0.3          & 4.0            & 1.9               & 0.010                   & 1.3                    \\
\hline
\hline
\end{tabular}
\\
\label{tab:rsfpar}
\end{table*}
%************************************************************************************%

The resulting models and the extracted data for the two adopted NLD normalizations are
shown in Fig.~\ref{fig:rsfmodels} (the models corresponding to the best fit for the lower and
upper normalization limits are not shown).
%---------------------------------------------------%
%---------------------------------------------------%
We see from Fig.~\ref{fig:rsfmodels} that the shape of the SLO model does not fit the
extracted RSF very well; in particular the low-energy part is very different in shape. 
The enhancement
of the RSF data at low $\gamma$-ray energies may be explained by the spin distribution of the initial levels
as described previously.

\subsection{Temperature dependence of the strength function}

\begin{figure}[bt]
 \begin{center}
 \includegraphics[clip,width=\columnwidth]{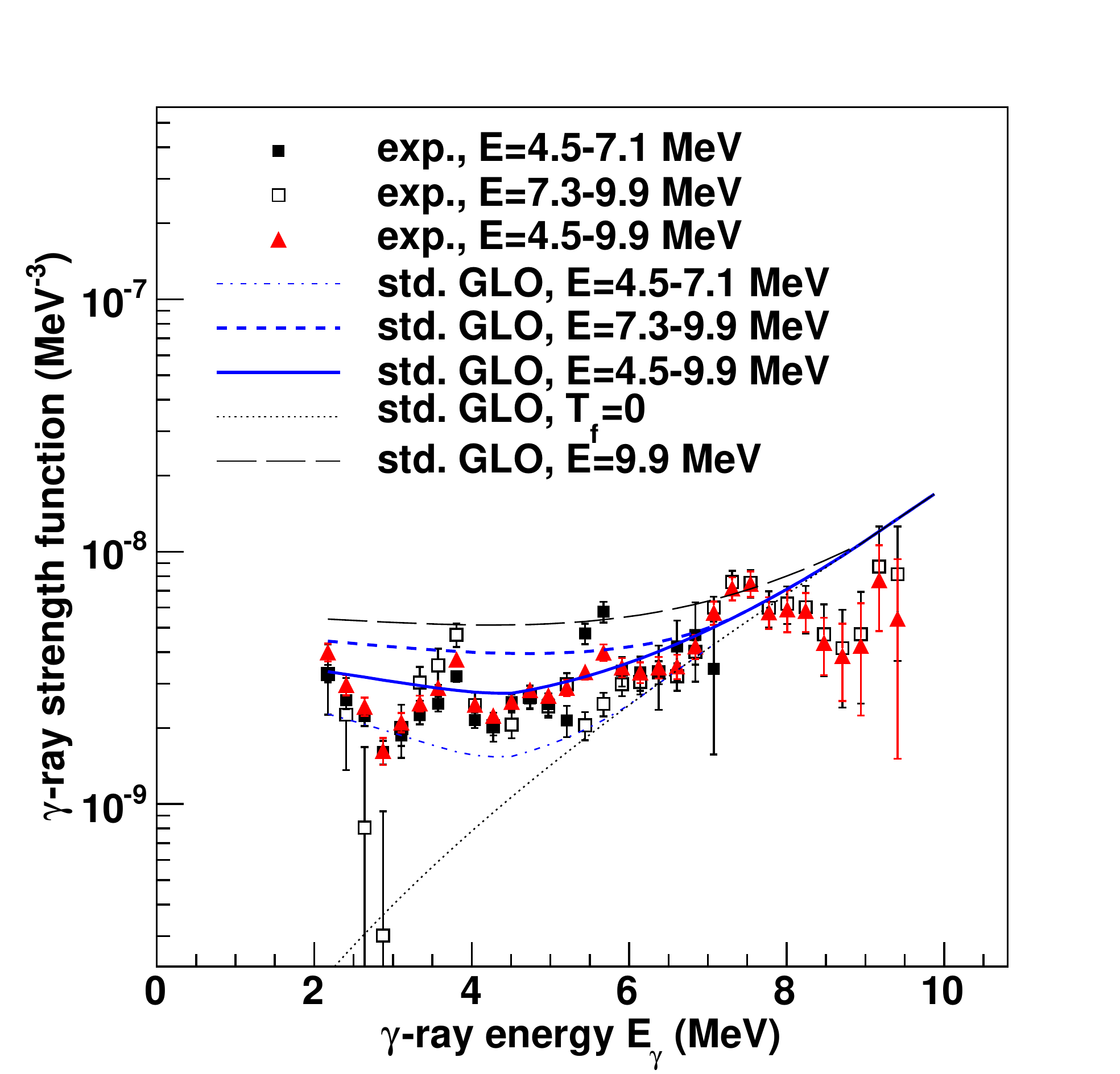}
%\vskip 2cm
 \caption{(Color online) Radiative strength functions 
	extracted for two different excitation-energy regions as compared to 
	the one from the total excitation-energy region under consideration (all cases are normalized to the CT level density). 
	The standard GLO model is also shown for 
	these ranges of excitation energy and the extreme cases $T_f=0$ and $E=9.9$ MeV.}
 \label{fig:test_T}
 \end{center}
 \end{figure}

As discussed already, we rely on the Brink hypothesis when extracting the NLD and RSF. We would like to investigate if
this hypothesis is reasonable, and 
we have therefore extracted the RSF for two different excitation-energy regions, 
$4.5 \leq E \leq 7.1$ MeV and $7.3 \leq E \leq 9.9$ MeV (see Fig.~\ref{fig:test_T}
for the result using the CT normalization).
We observe that there are rather large, local fluctuations, e.g. at $E_\gamma \approx 5.5$ MeV, that makes it hard 
to draw any firm conclusion whether the extracted RSF is dependent on excitation energy or not. 
As already mentioned, we expect large differences due to Porter-Thomas fluctuations in the decay strength of this nucleus.
However, the gross features seem to be quite similar for the two excitation-energy ranges. 
%In particular, for the interval $2.2 \leq E_\gamma \leq 5.2$ MeV,
%the deviation between the RSFs from the upper and lower regions is typically less than 40\%.
%---------------------------------------------------%
%---------------------------------------------------%
 
To get a better understanding on the possible temperature dependence, 
we have also used the standard GLO model with a varying temperature corresponding to the accessible final excitation energies
of the two ranges. The temperature is estimated by $T_f \propto \sqrt{E_f}$, and 
the models displayed in Fig.~\ref{fig:test_T} represent the average RSF within the three excitation-energy 
regions $4.5 \leq E \leq 7.1$ MeV, $7.3 \leq E \leq 9.9$ MeV, and $4.5 \leq E \leq 9.9$ MeV. 
The maximum final temperature reached is in the range 0.50--1.14 MeV 
for initial excitation energies between $4.5-9.9$ MeV and with
$E_\gamma^{\mathrm{min}} = 2.2$ MeV.

We observe that for $\gamma$ energies between $2.2 - 5.2$ MeV, the upper range can give more than a factor of two 
larger 
$\gamma$ strength in the GLO model than the lower range. 
The largest experimental fluctuations are also of the order of a factor two, 
but on average, the experimental data seem not to have such a strong dependence on the final excitation energy
as the standard GLO model would imply.

Finally, we have calculated the GLO model for the direct decay to the ground state ($T_f = 0$ MeV) and 
for the highest temperatures reached, namely for initial excitation energy $E=9.9$ MeV. These cases are also 
displayed in Fig.~\ref{fig:test_T}. It is clear that the zero-temperature calculation is not similar to 
the experimental data below $E_\gamma \approx 6$ MeV. In fact, the experimental data lie in between the 
two extreme cases. As argued above, the overall shape of the RSF seem to be rather similar for the 
two excitation-energy ranges. Thus, it is probably quite reasonable
to apply a constant temperature in the GLO model in accordance with the Brink hypothesis. 
We note however that the GLO model averaged over the total excitation-energy range covered 
(blue solid line in Fig.~\ref{fig:test_T}) agrees rather well with the average shape of the experimental data.

\section{Reproduction of experimental primary $\gamma$-ray spectra}
\label{sec:reprod}
%STEPH: inclusion of a new introductory paragraph
As described in the previous section, both the NLD and RSF deduced from the present 
experiment are affected by severe uncertainties associated with the normalization procedure 
that in the case of $^{44}$Ti cannot be reliably constrained by additional experimental data. 
For this reason, the different NLD and RSF models are now directly tested on the primary $\gamma$-ray spectra, 
which in turn correspond to the fundamental quantity entering the description of the radiative decay in reaction models.

%In order to check that the models applied are consistent with the experimental
%primary $\gamma$-ray spectra, we have compared calculated spectra with the 
%experimental ones. 
There are 24 experimental spectra for excitation 
energies $4.5 \leq E \leq 9.9$ MeV with bin size 233.4 keV. Due to the poor statistics, 
we have compared the average of two bins as in Fig.~\ref{fig:work}. For the level density,
we have applied the known levels for $E < 3.7$ MeV, and either the CT model or the GHK calculation
above this energy. The calculated spectra are scaled to get the best possible agreement with the 
experimental spectra, which are normalized such that $\sum_{E_{\gamma}=E_{\gamma}^{\mathrm{min}}}^{E} P(E, E_{\gamma}) = 1$.

We have also calculated the primary $\gamma$ spectra using the standard GLO model with a variable temperature in 
combination with the two level density models. 
For each initial excitation energy ($4.5 \leq E \leq 9.9$ MeV), we have used the
GLO model with $T_f \propto \sqrt{E_f}$ and made an average of all these RSFs for the 
whole excitation-energy region that is used for the analysis.

Finally, we have applied the SLO model in combination with the two NLD models.

\begin{figure*}[hbt]
 \begin{center}
 \includegraphics[clip,totalheight=11cm]{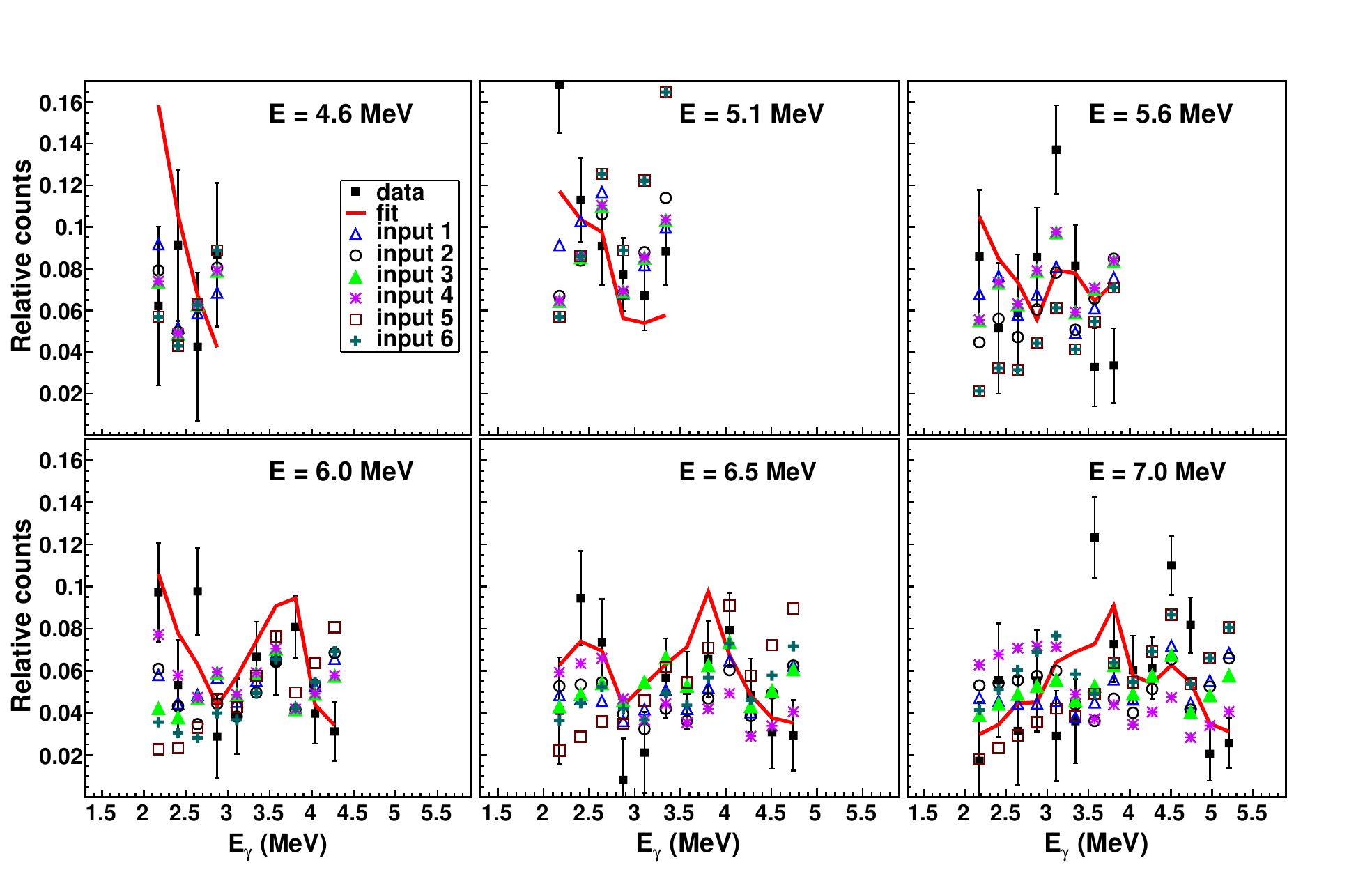}
%\vskip 2cm
 \caption{(Color online) Experimental primary $\gamma$ spectra (filled squares) and those obtained 
 from multiplying the extracted $\rho$ 
 and $\cal{T}$ functions (red line) for $4.6 \leq E \leq 7.0$ MeV. These are compared with  
 calculated spectra using the six inputs as described in the text. The experimental 
 and calculated spectra are given for excitation-energy bins of 466.8 keV.}
 \label{fig:work2}
 \end{center}
 \end{figure*}

\begin{figure*}[hbt]
 \begin{center}
 \includegraphics[clip,totalheight=11cm]{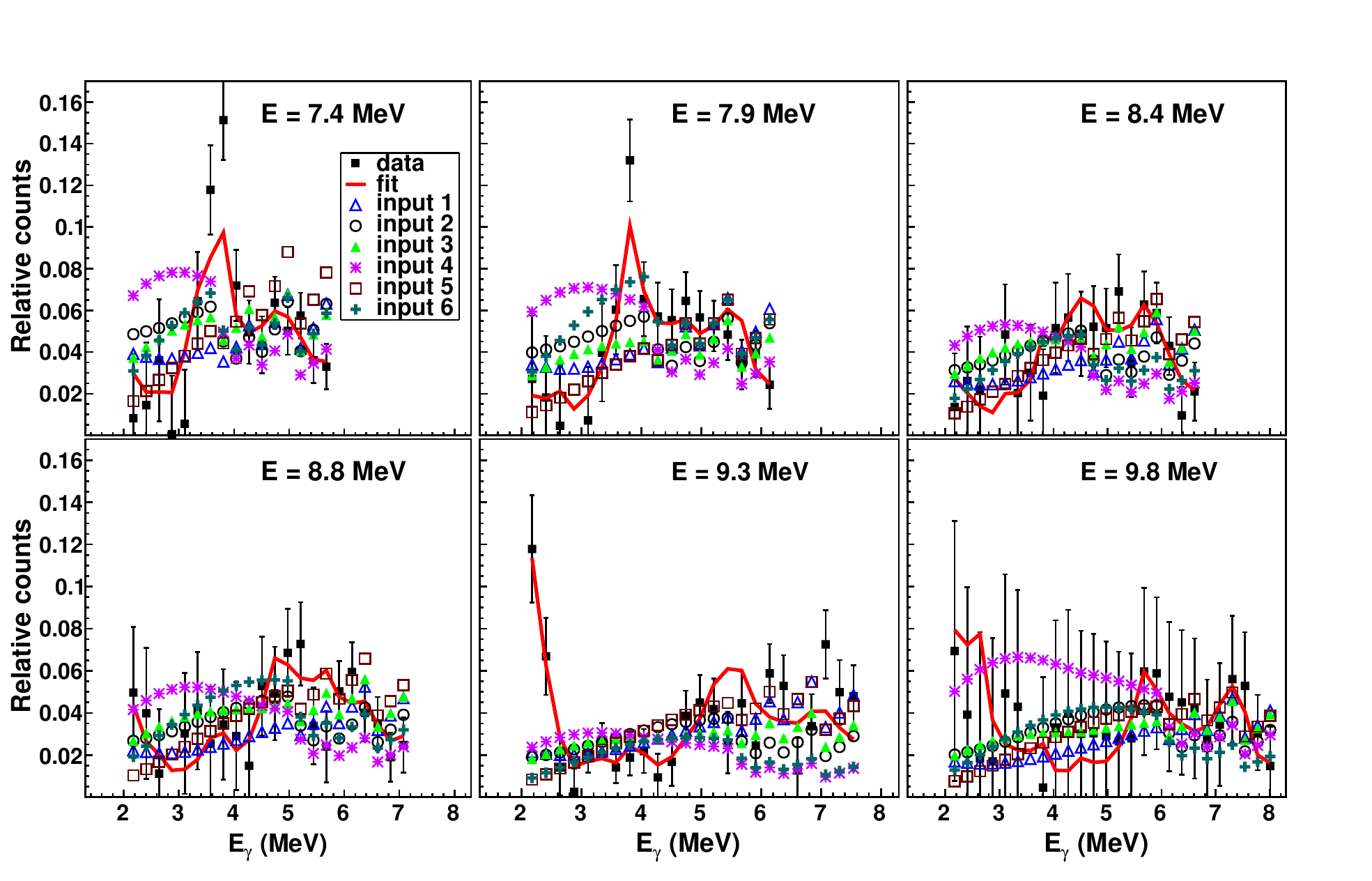}
%\vskip 2cm
 \caption{(Color online) Same as Fig.~\ref{fig:work2} for $7.4 \leq E \leq 9.8$ MeV.}
 \label{fig:work3}
 \end{center}
 \end{figure*}

The six model combinations shown in Figs.~\ref{fig:work2}--\ref{fig:work3} thus correspond to:
\begin{itemize}
\item input 1: the CT level density and the fitted GLO model with constant temperature.
\item input 2: the GHK level density and the fitted GLO model with constant temperature.
\item input 3: the CT level density and standard GLO model averaged over the experimental excitation-energy range 
	(blue, solid line in Fig.~\ref{fig:test_T}).
\item input 4: GHK level density and standard GLO model averaged over the experimental excitation-energy range (blue, solid line in
	Fig.~\ref{fig:test_T}).
\item input 5: the CT level density and the SLO model.
\item input 6: the GHK level density and the SLO model.
\end{itemize}
%---------------------------------------------------%
%---------------------------------------------------%
%---------------------------------------------------%
%---------------------------------------------------%

We find that for the lower excitation energies ($E \leq 7.0$ MeV), all the input models give a rather good 
reproduction of the spectra,
and also they give relatively similar results. However, for the higher excitation energies ($ 7.4 \leq E \leq 9.8$ MeV), there are
clear deviations for the different inputs, and input 4 gives a significantly worse %$\chi^{2}$ 
fit than the others. 
This is not so surprising, considering that the slope of the standard GLO model is very different from the RSF model fitted 
to our data using the GHK level density. It is seen in Fig.~\ref{fig:work3}
that this mismatch in slope leads to a wrong overall shape of the primary
spectra (overestimating the low-energy part and underestimating the high-energy part).
The models that are consistent in slope behave much better. 

In general, for the high-energy region, input 2 and input 3 give the best reproduction of the experimental spectra out of the 
six model combinations. The SLO model gives also rather good results, in particular in combination with the GHK level density. 
This could be an indication that the overall shape of this model might be 
correct, although the absolute value is probably too large (as seen in Fig.~\ref{fig:rsfmodels}). 
However, in combination with the CT level density the spectra calculated with this model 
underestimate the intensity of the low-energy $\gamma$ rays
compared to the measured ones. 

Naturally, using the extracted $\rho$ and
$\mathcal{T}$ data points give a much better fit than any model, as structures especially in the level
density are lost when using the smooth models. 
%Thus, for the optimal description of the measured primary spectra, the data points should be applied.

\section{Capture cross section and Maxwellian-averaged rate}
\label{sec:cross}

The model combinations described in Sec.~\ref{sec:reprod} can further be applied for estimating the 
$^{40}$Ca($\alpha,\gamma$)$^{44}$Ti cross section and the corresponding reaction rate.
We have used the code TALYS~\cite{TALYS} for the cross-section and reaction-rate calculations.
In all calculations we have used the $\alpha$ optical-model potential (OMP) of McFadden and Satchler~\cite{McFadden}.
We have also tested other $\alpha$-OMPs such as the one developed by Demetriou, Grama and Goriely~\cite{demetriou};
however, it turns out that in this case, the results are rather insensitive to the choice of $\alpha$-OMP. 
%and an incoming $\alpha$ energy of $1-10$ MeV.

The model sets applied for the cross-section and reaction-rate calculations are:
\begin{itemize}
\item input 1: the CT level density and the corresponding fitted GLO model.
\item input 2: the GHK level density and the corresponding fitted GLO model.
\item input 5: the CT level density and the SLO model.
\item input 6: the GHK level density and the SLO model.
\item input 7: the CT level density and the standard GLO model.
\item input 8: the GHK level density and the standard GLO model.
\item input 9: the CT level density and the corresponding fitted GLO model for the upper normalization limit
	($\left<\Gamma_{\gamma0}\right> = 1800$ meV).
\item input 10: the GHK level density and the corresponding fitted GLO model for the lower normalization limit
	($\left<\Gamma_{\gamma0}\right> = 600$ meV).
\end{itemize}

For the standard GLO model, we have followed the prescription used in TALYS, where the temperature of the final states 
is calculated from the relation~\cite{RIPL2}
\begin{equation}
T_f = \sqrt{(E_i - \Delta - E_\gamma)/a}.
\end{equation}
Here, $E_i$ is the initial excitation energy in the compound nucleus, $\Delta$ is a pairing correction, and $a$ is
the level density parameter. There is thus a specific GLO RSF for each initial excitation energy (no averaging over a large 
excitation-energy window as was done with inputs 3 and 4 in
Sec.~\ref{sec:reprod}). 

The Maxwellian-averaged astrophysical reaction rate $N_A \left< \sigma v \right> (T)$ for a temperature $T$ is given by
\begin{align}
N_A & \left< \sigma v \right> (T) =  \left( \frac{8}{\pi m} \right)^{1/2} \frac{N_A}{(kT)^{3/2}G(T)}  \times \\ \nonumber
	& \int_{0}^{\infty}
	\sum_{\mu} \frac{(2I^{\mu}+1)}{(2I^{0}+1)}\sigma^{\mu}(E) E \exp \left[ -\frac{(E+E_{x}^{\mu})}{kT} \right] dE,
\label{eq:maxwell}
\end{align}
where $m$ is the reduced mass of the initial system of projectile (here the $\alpha$ particle) and target nucleus (here $^{40}$Ca), 
$k$ is the Boltzmann constant, $N_A$ is Avogadro's number, $E$ is the relative energy of the target and projectile, 
$I^{\mu},E_{x}^{\mu}$ are the spin and excitation energy for the excited states labeled $\mu$, and $\sigma^{\mu}$ is the 
reaction cross section. Further, the temperature-dependent, normalized partition function $G(T)$ reads
\begin{equation}
G(T) = \sum_{\mu} (2I^{\mu}+1)/(2I^{0}+1) \exp (-E_{x}^{\mu}/kT).
\end{equation}
See the TALYS documentation for more details~\cite{TALYS}. 

For the above expressions, local thermodynamic equilibrium is assumed in the astrophysical environment, so that 
the energies of both the targets and projectiles as well as their relative energies $E$ obey Maxwell-Boltzmann 
distributions corresponding to the temperature $T$ at that location. Also the relative populations of the 
various nuclear levels with spin and excitation energy $I^{\mu},E_{x}^{\mu}$ obey a Maxwell-Boltzmann distribution.

\subsection{Uncertainties in the calculations}
The key ingredients in the rate calculations (NLD, RSF, and $\alpha$-OMP) all contribute to the uncertainties but
in different temperature regions. 
By varying the $\alpha$-OMP, the rate will change most at low temperatures (below $\approx 1.5 \cdot 10^{9}$ K),
while the NLD and RSF both have a significant impact on the rate at higher temperatures (above $\approx 1.5 \cdot 10^{9}$ K).

The $^{40}$Ca($\alpha,\gamma$) reaction only populates states with total isospin zero in 
$^{44}$Ti, and dipole transitions with no change in total isospin are suppressed in self-conjugate $N=Z$ nuclei \cite{Holmes}. 
Since complete
isospin mixing is assumed in the determination of the RSF, a corrective factor must be included before applying 
the extracted RSFs in the cross section and reaction rate calculations. For $\alpha$-captures, a standard 
prescription is to divide the RSF by a constant factor $f_{\mathrm{iso}}$ of typically 5 to 8 \cite{Nassar06,Holmes}, 
giving a reduced RSF $f'$ of $f' = f/f_{\mathrm{iso}}$. 

However, it is not obvious how the value of this correction factor should be determined, as it is related to the
degree of isospin mixing (for a large degree of mixing the correction factor should be small and vice versa). Also,
high-energy $\gamma$ rays decay to low-lying states, where the isospin mixing might be small, while low-energy
$\gamma$ rays decay in the quasi-continuum, where one expects a large degree of mixing. Therefore, one could in principle
expect that $f_{\mathrm{iso}}$ would vary as a function of $\gamma$ energy. To estimate such a 
function is however a very complicated task and beyond the scope of the present work. We will therefore assume a constant
$f_{\mathrm{iso}}$, although this is probably quite crude.

Another source of uncertainty is the value of $\left<\Gamma_{\gamma0}\right>$. Since this value is not known
experimentally, the uncertainty in this quantity is correlated to the uncertainty in $f_{\mathrm{iso}}$. This is because 
one can obtain basically the same cross section and reaction rate for a range of $\left<\Gamma_{\gamma0}\right>$ values 
by adjusting $f_{\mathrm{iso}}$ correspondingly (a small value of $\left<\Gamma_{\gamma0}\right>$ in combination with a small
$f_{\mathrm{iso}}$ and vice versa). 

We have calculated the integral cross section of the ($\alpha,\gamma$)$^{44}$Ti reaction for incoming $\alpha$
energies between $2.1-4.2$ MeV.
This corresponds to excitation energies in $^{44}$Ti in the range $E=7.2-9.3$ MeV, which is the most relevant region for
astrophysics.
In the calculations,
we have adopted a constant $f_{\mathrm{iso}} = 5$, and 
we have also tested a larger correction factor of $f_{\mathrm{iso}} = 8$. In addition, we have considered
the assumed uncertainty of 50\% in the estimated value of $\left<\Gamma_{\gamma0}\right>$ combined with the 
uncertainty in slope (either the CT or the GHK level density, which represent 
the extremes in slope of the extracted RSF). 
The results for all the considered
inputs are given in Tab.~\ref{tab:cross}. 
%************************************************************************************%
\begin{table}[htb]
\caption{Integral cross sections for the various model combinations.} 
\begin{tabular}{lrr}
\hline
\hline
Model combination					 &\multicolumn{2}{c}{$\sigma_{\mathrm{ave}}$ ($\mu$b)}  \\
    &  $f_{\mathrm{iso}}=5$  &  $f_{\mathrm{iso}}=8$		\\
\hline
input 1             & 7.1    & 4.9   \\
input 2             & 6.9    & 4.9   \\
input 5             & 17.8    & 13.2  \\
input 6             & 19.2    & 14.5  \\
input 7             & 5.6    & 3.8  \\
input 8             & 6.9    & 4.9  \\
input 9             & 9.7    & 6.9  \\
input 10            & 5.0    & 3.4   \\
\hline
\hline
\end{tabular}
\\
\label{tab:cross}
\end{table}
%************************************************************************************%

We see from Tab.~\ref{tab:cross} that changing the isospin correction factor leads to a change in the calculated
cross section of typically 2--3 $\mu$b, except for the SLO model where different values of $f_{\mathrm{iso}}$ 
give up to $\approx 5$ $\mu$b change in $\sigma_{\mathrm{ave}}$. The uncertainty in $\left<\Gamma_{\gamma0}\right>$
and slope will also give a change in $\sigma_{\mathrm{ave}}$ of up to $\approx 5$ $\mu$b. 

We therefore conclude that the absolute value of the reaction cross section 
of the $\alpha$ capture on $^{40}$Ca is highly uncertain. Also, intrinsically very different models of the level density and the RSF 
may yield practically the same cross section by adjusting $f_{\mathrm{iso}}$ and/or $\left<\Gamma_{\gamma0}\right>$.
Thus, we find that although our data may put a constraint on the functional form of the NLD and RSF, and in particular the correlated
slope of the two quantities, other data are needed in order to further constrain the cross section and reaction rate. This
will be addressed in the following.

\subsection{Comparison with other data}
We have compared our calculations with two recent data sets, one from Nassar \textit{et al.}~\cite{Nassar06} and one
from Vockenhuber \textit{et al.}~\cite{vockenhuber07}. In the following discussion we have used $f_{\mathrm{iso}} = 5$,
unless stated otherwise.

Nassar \textit{et al.}~\cite{Nassar06} performed an integral measurement on the $^{40}$Ca($\alpha,\gamma$)$^{44}$Ti
cross section corresponding to an energy window of $E_\alpha = 2.1-4.2$ MeV for the incoming $\alpha$ particles. The 
originally estimated energy-averaged cross section was $\sigma^{\mathrm{exp}}_{\mathrm{ave}} = 8.0(11)$ $\mu$b; 
however, it was pointed
out by Vockenhuber \textit{et al.}~\cite{vockenhuber07} that due to an overestimate of the $^{40}$Ca stopping power, the 
cross section should be reduced by about 10\%: $\sigma^{\mathrm{exp}}_{\mathrm{ave}} \simeq 7.0(10)$ $\mu$b. 

Using the model combinations input 1 and input 2, we get 
$\sigma_{\mathrm{ave}} = 7.1$ and 6.9 $\mu$b for the CT and GHK normalization, respectively (see Tab.~\ref{tab:cross}). 
This is in excellent agreement with
the Nassar results. Note that if we exclude the low-energy enhancement in input 1 and 2, 
$\sigma_{\mathrm{ave}}$ yields 6.6 and 6.7 $\mu$b for the CT and GHK normalization, respectively. Thus,
the effect of the small low-energy enhancement is 
not significant compared to neither the current experimental uncertainty on the integral cross section, 
nor the uncertainties in the calculations
due to the isospin suppression factor and the unknown absolute normalization of the RSF.

Further, if we apply the standard GLO model in the calculations, we obtain
$\sigma_{\mathrm{ave}} = 5.6$ and 6.9 $\mu$b using the CT and GHK level density (input 7 and input 8), respectively. The former
is slightly smaller than the lower experimental limit on $\sigma^{\mathrm{exp}}_{\mathrm{ave}}$. The latter agrees very well 
with the Nassar data; however, we noted in Sec.~\ref{sec:reprod} that the GHK level density in combination with the averaged,
standard GLO did not reproduce our experimental spectra above $E\approx 7$ MeV. Therefore, input 8 is probably not correct.

The calculations with the SLO model are many standard deviations too large compared to the Nassar data, even
for $f_{\mathrm{iso}} = 8$. Thus, we will not use inputs 5 and 6 further. We also see that the upper normalization
limit, input 9, is only acceptable for $f_{\mathrm{iso}} = 8$. On the other hand, the lower normalization limit
(input 10) seems to be a bit too low for $f_{\mathrm{iso}} = 5$ and even more so for $f_{\mathrm{iso}} = 8$. 

\begin{figure}[bt]
 \begin{center}
 \includegraphics[clip,width=\columnwidth]{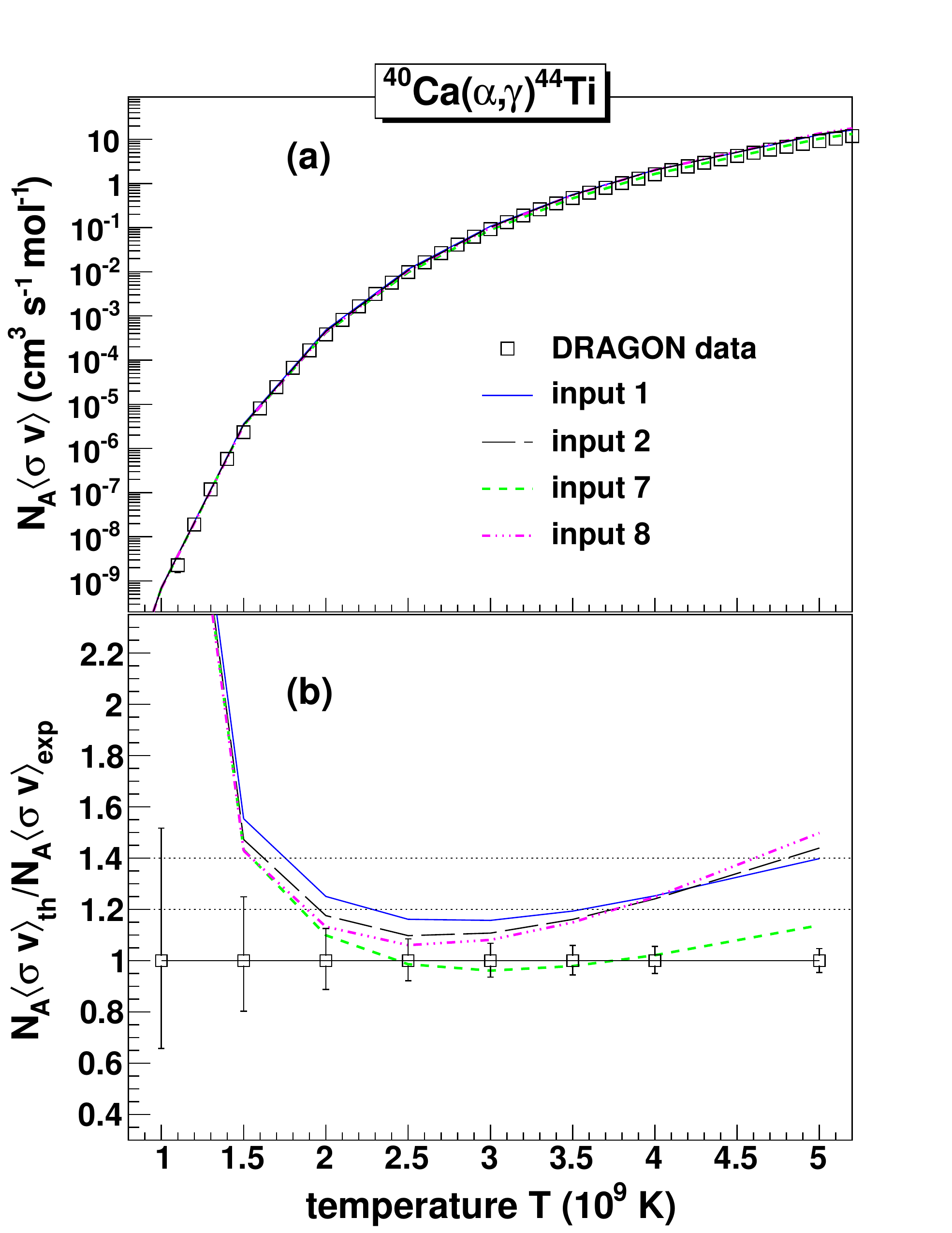}
%\vskip 2cm
 \caption{(Color online) (a) Data on the Maxwellian-averaged reaction rate 
	for the $^{40}$Ca($\alpha,\gamma$) reaction 
	from Ref.~\cite{vockenhuber07} are compared to calculations with 
	inputs 1, 2, 7 and 8. (b) 
	Ratio of the calculated reaction rates  
	and experimental data. The upper dotted line indicates
	a 40\% deviation from the experimental data, while the lower dotted line indicates a 20\% deviation.}
 \label{fig:rates}
 \end{center}
 \end{figure}

With the inputs 1,2,7 and 8, we have estimated the reaction rate according to Eq.~(10), and compared to the 
DRAGON data of Vockenhuber \textit{et al.}~\cite{vockenhuber07}. 
The results are shown 
in the upper panel of Fig.~\ref{fig:rates}. 
%---------------------------------------------------%
%---------------------------------------------------%
As seen here, all these four inputs give a reasonable reproduction of the measured 
rates, and they can hardly be distinguished from each other
over the entire temperature range. We also note that the cross section range spans over 
ten orders of magnitude, and it is quite impressive how well the calculated rates follow the overall 
functional form of the data. 

However, we observe that some models overestimate the rate somewhat, 
especially at higher temperatures. Now, the experimental rate could be too low for temperatures above
$T_9 \approx 4$ MeV because of missing resonances (only six resonances with measured resonance 
strengths are known for excitation energies at and above 9.3 MeV~\cite{vockenhuber07}). We note however that
the rate at these high temperatures is not important for the final $^{44}$Ti mass yield~\cite{vockenhuber07}.

The combination of CT 
level density and standard GLO (input 7) gives a better reproduction of the DRAGON data at high temperatures than the others. 
However,
this combination gave a too small average cross section as compared to the Nassar result. We therefore confirm the slight 
inconsistency between the two measurements, as discussed in Ref.~\cite{vockenhuber07}. As already mentioned, 
we observed that the combination
of the GHK level
density and the averaged, standard GLO model (input 4) was not able to give a reasonable description of our primary $\gamma$
spectra for initial excitation energies above $\approx 7$ MeV. Therefore, input 8 cannot be recommended on 
the basis of the present data. 

In order to get a clearer picture of the deviation of the calculated rates vs. the DRAGON data, 
the ratio of the calculated and experimental rate is shown in the lower panel of Fig.~\ref{fig:rates}. Here, it is seen that
input 7 gives the best fit to the DRAGON data for all temperatures above 
$T_9 \approx 1.7$ ($T_9 = 10^9$ K). The large overestimate of the rate for $T_9 < 1.5$ is common for all model
predictions and is due to problems with the $\alpha$ OMP at very low energies.

It is seen from the lower panel in Fig.~\ref{fig:rates} that all the calculated rates lie within 40\% of the DRAGON data for 
$1.8 \leq T_9 \leq 4.5$, and for the 20\% upper limit in the range $2.2 \leq T_9 \leq 3.7$.
However, only input 7 (CT level density and standard GLO) is within the experimental 
error bars for $1.8 \leq T_9 \leq 4.5$. 

It should be noted that by using the lower normalization limit
for the RSF (blue, dashed-dotted line in Fig.~\ref{fig:rsfs}), an excellent agreement with the DRAGON data is 
obtained. However, as stated previously, a good reproduction of the DRAGON data implies a too low integral cross section 
(in this case $\sigma_{\mathrm{ave}} = 5.0$ $\mu$b) as compared to the Nassar data. On the other hand, using the
upper normalization limit for the RSF (dashed line in Fig.~\ref{fig:rsfs}), it is necessary to use the larger value 
of $f_{\mathrm{iso}} = 8$ to obtain a reasonable agreement with the data (see Tab.~\ref{tab:cross}). 

To summarize, we have found that the model combinations input 1, input 2, and input 8 are in excellent agreement with the 
Nassar cross-section measurement. However, input 8 (GHK level density and standard GLO) is not in accordance with our primary
$\gamma$ spectra for the relevant excitation-energy region ($E=7.2-9.3$ MeV) and should therefore not be used. 
The DRAGON data are best described with input 7
(CT level density and standard GLO) and the lower normalization limit of our RSF data (input 10). 
We also see that the small enhancement in the RSF at low $E_\gamma$ is not very important for the integrated cross section
or the rate, as it gives a contribution of maximum 0.5 $\mu$b. 

%% STEPH: I would skip this paragraph for the time being.
%The precision needed for astrophysical calculations of the $^{44}$Ti mass
%fraction is $\approx 20$\%, and this is thus the case for all models in the temperature range $2.2 \leq T_9 \leq 3.7$.
%From the sensitivity study in Ref.~\cite{vockenhuber07}, it is found that the final mass fraction yield of 
%$^{44}$Ti does not depend on the reaction rate for temperatures above $T_9=4.3$ due to the equilibrium
%condition in the stellar medium. For the interval  $2.8 < T_9 < 4.3$, a weak dependence is observed, while for 
%temperatures in the range $1.0 < T_9 < 2.8$, the final yield shows a strong dependence on the reaction rate. 
%This is thus the most crucial interval for the obtained $^{44}$Ti mass fraction. 

%\textbf {Stephane: is there any point in trying to estimate the mass 
%fraction with our calculated rates as input, or not? } \newline
% STEPH: I don't believe there is a need, since we do not have a specifically different cross section !

\section{Summary}
\label{sec:con}
Particle-$\gamma$ coincidence data of the $^{46}$Ti($p,t\gamma$)$^{44}$Ti reaction have been measured 
at OCL. By use of the Oslo method, primary $\gamma$-ray spectra have been extracted for initial excitation
energies in the range $E=4.5-9.9$ MeV. From these spectra, the functional form of the level density and
the RSF have been determined. 

%Litt mer om NLD og RSF + temp og Ex

We have shown that a consistent normalization of the NLD and RSF is necessary in order to obtain a 
reasonable reproduction of the primary spectra. Also, the RSF seems to be independent of excitation energy
and thus of the temperature, in accordance with the Brink hypothesis. However, the GLO model with 
a variable temperature, and averaged over the excitation-energy region, gives a rather good description of 
the overall shape of our RSF data when normalized to the CT level density. 

Using input models consistent with our data give an excellent reproduction of the 
Nassar integral measurement of the $^{40}$Ca($\alpha,\gamma$) reaction cross section. Also the DRAGON
data of the reaction rate are rather well reproduced. However, we note that there is a discrepancy between
the two datasets, so that it is not possible to use one model combination to obtain optimal agreement for
both measurements simultaneously. Nevertheless, on the basis of our present data, it is clear that certain
model combinations are not acceptable, although they might give reasonable results compared to the cross-section
data.

\begin{acknowledgments}
Funding of this research from the Research Council of Norway, project grant no. 180663, is gratefully acknowledged. We would 
like to give special thanks to 
E.~A.~Olsen and J.~Wikne for providing excellent experimental conditions. 
A.~C.~L. would like to thank M.~Hjort-Jensen and M.~Krti\u{c}ka for inspiring and enlightening discussions.
\end{acknowledgments}

%\vfill
\end{document}